\documentclass[aip,jcp,amsmath,amssymb,preprint]{revtex4-1}
\usepackage{graphicx}

\usepackage{dcolumn}
\usepackage{bm}
\usepackage{xcolor}
\usepackage{amsthm}
\newtheorem*{theo}{Theorem}

\begin{document}
 
\preprint{AIP/JCP/}
\hfill {\small \today}

\begin{center}{\large\bf
On the large interelectronic distance
 behavior of the correlation factor 
 for explicitly correlated wave functions}
\vspace{3ex}

{\sc Micha\l~Lesiuk$^a$, \  Bogumil~Jeziorski, \ Robert Moszynski}
\vspace{1ex}

{\it Faculty of Chemistry, University of Warsaw, Pasteura 1, 02-093 Warsaw, Poland}
\vspace{4ex}

\end{center}

{\small
\centerline{Abstract}
\vspace{1ex}

In currently most popular explicitly correlated electronic structure
theories the dependence of the wave function on the interelectronic distance
$r_{ij}$ is built via the correlation factor $f(r_{ij})$.  While
the short-distance behavior of this factor is well understood,
little is known about the form of $f(r_{ij})$ at large $r_{ij}$.
In this work we investigate the optimal form of $f(r_{12})$ on the
example of the helium atom and helium-like ions and several
well-motivated models of the wave function.
Using the Rayleigh-Ritz variational principle we derive
a differential equation for $f(r_{12})$ and solve it using
numerical propagation or analytic asymptotic expansion techniques.
We found that for every 
model under consideration, $f(r_{12})$ behaves at large 
$r_{ij}$ as $r_{12}^\rho\,e^{B r_{12}}$
and obtained simple analytic expressions for the system dependent values of 
 $ \rho$ and $B$. 
For the ground state of the helium-like ions the value of $B$ is positive,
so that $f(r_{12})$ diverges as $r_{12}$ tends to infinity. The numerical
propagation confirms this result.
When the Hartree-Fock orbitals, multiplied by the correlation factor, are
expanded in terms of Slater functions $r^n e^{-\beta r}$, $n=1\ldots N$, the
numerical propagation reveals a minimum in $f(r_{12})$ with depth
increasing with $N$. 
For the lowest triplet state $B$ is negative. 
Employing our analytical findings, we propose a new ``range-separated''
form of the correlation factor with the short- and long-range $r_{12}$ 
regimes approximated by appropriate asymptotic formulas 
connected by a switching function. Exemplary calculations show that 
this new form of  $f(r_{12})$  
performs somewhat  better than the correlation factors used thus far 
in the standard R12 or F12 theories.
}

\vfill
\noindent{\small $^a$e-mail: lesiuk@tiger.chem.uw.edu.pl}


\newpage 
\section{\label{sec:intro}Introduction}

It is well known that the slow convergence of the standard, orbital based methods 
of the electronic structure theory  is due to 
the difficulties to model the exact wave function in the regions 
of the configurations space where  electrons 
are close to each other\cite{hattig12,kong12}.
It was shown by Kato\cite{kato57} and later elaborated by Pack and 
Byers-Brown\cite{pack66}, and Hoffman-Ostenhofs \emph{et al.}\cite{furnais05,hoffmann92} that in the vicinity of points
where the positions of two electrons coincide, the wave function 
behaves linearly in the interelectronic distance $r_{12}$. 
Such a behavior, referred often to as the \textit{cusp condition}, 
cannot be modeled by a finite expansion in terms of orbital 
products\cite{szalewicz10}. The solution to this 
problem is to include the interelectronic distance dependence 
directly into the wave function. This is the main idea of 
the so-called 
\textit{explicitly correlated methods} of the electronic structure 
theory\cite{hattig12,kong12,tenno12}. 
It should be noted however, that the explicit dependence on $r_{12}$
is advantageous even if the cusp condition is not fulfilled exactly
as in the Gaussian geminal\cite{szalewicz10,bukowski03} 
or the ECG\cite{rychlewski03,bubin13} (explicitly correlated Gaussian) 
approaches. This is due to the fact   
that the correlation hole, i.e., the decrease of the wave function 
amplitude when the electrons approach each other, is much easier to model 
with basis functions depending explicitly on $r_{12}$ than with 
the orbital products\cite{szalewicz10}.

The simplest way to make the wave function $r_{12}$ dependent
is to multiply some or all orbital products in its conventional 
configuration-interaction-type expansion by a {\it correlation factor}  
$f(r_{12})$. In this way all $r_{12}$ dependence is contracted 
in one function of single variable. The idea of the correlation factor
is very old one. It can be traced back to the late  1920's work 
of Slater\cite{slater28} and of Hylleraas\cite{hylleraas29a,hylleraas29b} 
who showed great effectiveness of including the linear $r_{12}$ 
term in the helium wave function. More than two decades later  
Jastrow\cite{jastrow55} proposed to use the correlation factor 
to construct a compact form of correlated wave function for 
an N-particle quantum system. The wave function form proposed 
by Jastrow became popular in the electronic structure 
theory as the guide function in diffusion-equation Monte-Carlo 
calculations\cite{luchow00,foulkes01}.

The concept of the correlation factor is now most widely used 
in the context of many-body perturbation theory\cite{kucharski86}
 (MBPT) and coupled cluster\cite{bartlett07} (CC) approach. 
It was first observed by Byron and 
Joachain\cite{byron66}, and later by Pan and King\cite{pan70,pan72},
Szalewicz and co-workers,\cite{chalas77,szalewicz79,szalewicz82,szalewicz83b,szalewicz84}, and Adamowicz and
Sadlej\cite{adamowicz77,adamowicz78a,adamowicz78b} that 
the pair functions appearing in the energy
expressions of the MBPT or CC theory can be very efficiently approximated when 
expanded in terms of explicitly correlated basis functions.
In the investigations of Refs. 
\cite{pan70}--
\cite{adamowicz78b} 
the dependence on the $r_{12}$ coordinate was introduced through 
the Gaussian factors, $\exp{(-\gamma_i r^2_{12})}$, with 
different $\gamma_i$  for different basis functions 
(Gaussian geminals). Thus, the pair functions 
were  not represented with a single, universal correlation factor.  
Massive optimizations of thousands of nonlinear parameters defining the Gaussian geminals
($\gamma_i$  and orbital exponents)  
made these calculations very time-consuming, limiting applications 
of this approach to very small systems like Be, Li$^-$, LiH, He$_2$, 
Ne, or H$_2$O\cite{bukowski99,przybytek09,patkowski07,wenzel86,bukowski95}. 

An important advance in the field of explicitly correlated MBPT/CC theory 
came with the seminal 1985 work of Kutzelnigg\cite{kutz85} and 
the subsequent development of the so-called R12 method by Kutzelnigg, 
Klopper and Noga\cite{klopper87,kutz91,noga92,noga94,klopper03}. 
In this work  a simple linear correlation factor $f(r_{12}) =r_{12}$
was used to multiply products of occupied Hartree-Fock (HF) 
orbitals $\phi_i$, $i=1,\ldots n$. 
The resulting set of explicitly correlated 
basis functions $f(r_{12})\phi_i\phi_j$, supplemented by products 
of all virtual orbitals, was then used to expand the pair functions 
of the MBPT/CC theory. 
The necessity to
calculate three and four-electron integrals, resulting from 
the Coulomb and exchange operators and the strong orthogonality projectors, 
is eliminated by suitable resolution of identity (RI) insertions. 
Kutzelnigg and Klopper introduced also some useful
approximations\cite{klopper87,kutz91} to the expression 
for the commutator of the Fock operator with $f(r_{12})$ 
which significantly simplified calculations. 
The practical implementation of the original R12 scheme was, 
however, not free from problems. 
Most importantly, in order to make the RI approximation accurate
enough the one-electron basis set used in calculations had   
to be very large. This constraint was alleviated  by 
Klopper and Samson\cite{klopper02} who introduced auxiliary 
basis sets for the RI approximation which are saturated 
independently from the size of the basis set that is used 
in the preceding Hartree-Fock calculations.
During the past two decades the R12 technology was 
progressively refined by the use of many tricks such 
as the density fitting\cite{manby03}, numerical 
quadratures\cite{tenno04a}, improvements in the RI
approximations\cite{tenno03,valeev04a}, or efficient 
parallel implementations.\cite{valeev00,valeev04b}
A generalizations to multi-reference 
configuration interaction problems (MRCI-R12) have 
been developed by Gdanitz\cite{gdanitz93,gdanitz98}.
One should also mention the work of 
Taylor and co-workers\cite{persson96,persson97,dahle01} 
who expanded the linear correlation factor $r_{12}$ 
as a combination of the Gaussian functions, 
and evaluated the necessary many-electron integrals   
analytically.  

Despite this progress, the results of R12 calculations 
using small basis sets were l not fully  satisfying. 
In particular, it was shown that the results of R12 
calculations with  a correlation-consistent polarized 
valence double-zeta (cc-pVDZ) basis set were of similar 
quality as ordinary orbital based  calculations 
with a triple-zeta cc-pVTZ basis set\cite{klopper02}. 
This is a rather small gain when compared to the accuracy 
improvement in calculations with the quintuple-zeta basis 
sets when the R12 method gives almost saturated results. 
In 2005 May and co-workers\cite{may05} reported a careful 
analysis of the errors in R12 theory at the second-order
M\o ller-Plesset (MP2-R12) level. They concluded that 
the most significant source of these errors are  defects
inherent in the R12 {\em Ansatz} and that it is essential
that $r_{12}$ is replaced by a more accurate correlation 
factor $f(r_{12})$. Actually, a  generalization 
of the R12 theory, referred to as
the F12 theory, allowing an arbitrary, nonlinear correlation 
factor $f(r_{12})$  was formulated by May and Manby\cite{may04}
already in 2004. In the same year Ten-no\cite{tenno04b}
proposed  the use of the exponential correlation factor 
$[1-\exp(-\gamma r_{12})]/\gamma$ 
(Slater-type geminal) and showed that it leads to much better 
results than the linear one. This launched   
rapid development of the F12 methods, which are now almost 
exclusively based on the application of the exponential 
correlation factor.\cite{kong12,tenno12}. 
 This correlation factor turned out to be effective not only in the 
conventional single-reference MBPT/CC theory but 
was also successfully applied to improve the basis set convergence 
of multireference methods: MRCI\cite{shiozaki11a,shiozaki11b}, 
multireference perturbation theory \cite{torheyden09,tenno07,shiozaki10},
multireference CC approach \cite{kedzuch11}, and even 
the multiconfiguration SCF procedure\cite{martinez10}. 

It is clear that the shape of the correlation factor is 
important for the high quality of the results. 
One may, thus, ask what is the optimal form of $f(r_{12})$ 
that is correct not only in the vicinity of the electrons 
coalescence points, 
but also at arbitrary distance between electrons. 
This question has been considered by Tew and Klopper\cite{tew05} 
who have investigated the shape of the correlation factor for the
helium atom and for helium-like ions and compared it 
with several simple analytic forms. These authors expanded $f(r_{12})$ as a polynomial in 
$r_{12}$ and determined its coefficients by minimizing the distance  
(in the Hilbert space) between the exact wave function
and its approximate form constructed using $f(r_{12})$. They found that 
the exponential correlation factor proposed by Ten-no\cite{tenno04b}
is  close to   optimal.  

It should be pointed out that the method used by Tew and Klopper\cite{tew05}
is not accurate at larger values of $r_{12}$ and does not give any 
information about the asymptotic   
behavior of $f(r_{12})$ at large $r_{12}$.
This is a consequence of the assumed polynomial form for $f(r_{12})$,
which prejudges the asymptotic behavior of $f(r_{12})$
and makes the obtained approximation to the optimal $f(r_{12})$  
less reliable at larger $r_{12}$. Moreover, the optimum  $f(r_{12})$ as 
defined by Tew and Klopper does not guarantee the minimum energy 
with respect to a variation of a fully flexible form of 
the correlation factor.    

In the present communication we propose an alternative method 
to determine the optimal form of $f(r_{12})$, which is free from the 
above drawbacks. We do not expand $f(r_{12})$ in a basis set
but derive a differential equation for $f(r_{12})$, 
resulting from the unconstrained minimization of the Rayleigh-Ritz 
energy functional. This differential equation can be solved by
a numerical propagation or using analytic, asymptotic expansion techniques. 
In this way the problems with the stability of the optimal $f(r_{12})$
at large $r_{12}$, experienced by Tew and Klopper\cite{tew05},
are avoided and we obtain a reliable information on the large $r_{12}$ 
behavior of $f(r_{12})$.  This information, combined with the well known
information about the short-range behavior of  $f(r_{12})$, 
gives us a possibility to propose a new form of the correlation 
factor which is correct at small and large values of $r_{12}$.
One may hope that the correlation factor  more adequate at large 
$r_{12}$ will make up for the lack of flexibility of the 
orbital basis to describe the long-range correlation and will reduce the 
basis-set requirements of  F12 calculations. 

The paper is organized as follows. In Sections \ref{subsec:1s1s} and \ref{subsec:1s2s} we analyze the  simplest models
of the correlated wave functions 
for the ground and the lowest triplet state of the helium atom and helium-like ions.  
In both cases, we establish differential equations for the correlation factor $f(r_{12})$ 
and solve them exactly in the large-$r_{12}$ domain. In Section \ref{subsec:gauss1s} we investigate another model for
the singlet ground state when the 1$s$ Slater orbital is replaced by 
a single Gaussian function.
In Section \ref{subsec:hff12} we move on to the case of
a self-consistent-field (SCF) determinant  multiplied by the correlation factor. 
In this case, we were not able to derive an explicit
differential equation but we present equations 
sufficient to determine the leading term of the asymptotic expansion for
$f(r_{12})$. In Section \ref{subsec:kutzf12} we report changes that occur when a set of excited state determinants is
added to the approximate wave functions considered previously. In Section \ref{sec:numerical} we propose a new
analytical form of the correlation factor and give results of simple numerical calculations, followed by a short
discussion. The paper ends with conclusions in Sec. \ref{sec:conclusions}.

In our work we use several special functions.  
The definition of these functions is the same as in 
Ref. \cite{stegun72}. Atomic units are used throughout the paper.

\section{\label{sec:theo}Theory}

\subsection{\label{subsec:1s1s}Correlated Slater orbitals. Singlet state.}

We first consider a very simple model, a particular case of the 
Slater-Jastrow wave function\cite{jastrow55,foulkes01} for  
helium-like ions:
\begin{align}
\label{psi1}
\Psi= \Psi_0(r_1,r_2)  f(r_{12}),
\end{align}
where $r_1$ and $r_2$ are the electron-nucleus distances, 
$r_{12}$ is the interelectronic distance, 
$ \Psi_0(r_1,r_2)= e^{-\alpha r_1}\, e^{-\alpha r_2}$ and  $f(r_{12})$ is the
correlation factor. The orbital exponent $\alpha$ is  left unfixed  -- 
it can be later optimized  without or with the correlation factor. 
We determine  $f(r_{12})$ by unconstrained minimization of 
the Rayleigh-Ritz  energy functional:
\begin{align}
\label{en1}
E[f]=\frac{\langle \Psi_0 f|\hat{H}|\Psi_0 f\rangle}
{\langle \Psi_0 f|\Psi_0f\rangle}.
\end{align}
The requirement that the functional derivative of $E[f]$ is zero,
\begin{align}
\label{varia}
\frac{\delta E}{\delta f(r_{12})}=0,
\end{align}
or equivalently that
\begin{align}
\label{variabj}
\frac{\partial E[f+\mu \delta f]}{\partial \mu} \Big|_{\mu=0}=0,
\end{align}
for every variation $\delta f$ of $f$, leads to a differential 
equation for $f(r_{12})$. 
This equation has a unique solution (up to a phase) 
if we assume that $f$ is regular at $r_{12}=0$ and 
that $\Psi=\Psi_0 f$ is square integrable.

To evaluate the functional derivative of Eq. (\ref{varia})
it is convenient to integrate over Euler angles first 
and perform the integral over $r_{12}$ at the end. This 
can be done by means of the formula:
\begin{align}
\label{inter}
\int\! \!\int \mathcal{F}(r_1,r_2,r_{12})d\textbf{r}_1 d\textbf{r}_2\;
=8\pi^2 \int_0^\infty\!\!\int_0^\infty \!\!
\int_{|r_1-r|}^{r_1+r} \; r_1 r_2 r\; \mathcal{F}(r_1,r_2,r)\, dr_2\, dr_1\, dr,
\end{align}
where $\mathcal{F}(r_1,r_2,r_{12})$ is any function 
for which the integral on the left exists. 
For states of $S^e$  symmetry  and  wave functions  expressed through interparticle distances  
$r_1$, $r_2$, and $r_{12}\equiv r$  the Hamiltonian can be taken in the form
\begin{align}
\label{hamilts}
\begin{split}
\hat{H}=-\frac{1}{2}\left(1+\mathcal{P}_{12}\right)
\left[
\frac{\partial^2}{\partial r_1^2}+
\frac{2}{r_1}\frac{\partial}{\partial r_1}+
 \frac{r^2+r_1^2-r_2^2}{r r_1 }\frac{\partial^2}{\partial r_1 \partial r}+
\frac{2Z}{r_1}\right]  - \frac{\partial^2}{\partial r^2} -
\frac{2}{r}\frac{\partial}{\partial r}
+\frac{1}{r},
\end{split}
\end{align}
where $\mathcal{P}_{12}$ denotes permutation of the indices $1$ and $2$, and $Z$ is the nuclear charge.
In Eq. (\ref{hamilts}) and in the following text we  
denote $r_{12}$ by $r$ to make equations  more transparent
and more compact. 
 Recently,
Pestka\cite{pestka08} presented generalizations of this Hamiltonian 
valid for two-electron states of arbitrary angular momentum. 
His results can be used to extend our approach to states
of higher angular momenta.

Evaluating the l.h.s. of Eq. (\ref{variabj}) with the help of 
Eq. (\ref{inter}) and assuming that it vanishes  for every variation 
 $\delta f$ one obtains the following equation for $f$
\begin{align}
\label{symimpl}
\int_0^\infty \!\!\int_{|r_1-r |}^{r_1+r } r_1 r_2  e^{-\alpha(r_1+r_2)}
\left(\hat{H}-E\right) e^{-\alpha(r_1+r_2)} f(r ) dr_2 dr_1  =0.
\end{align}
To obtain the explicit form of this equation we have 
to perform integration over the variables $r_1$ and $r_2$. 
Using Eq. (\ref{hamilts}) and the integral formulas from Appendix A 
one finds
\begin{align}
\label{dif1s1s}
\begin{split}
&\left[
-3
+3\left(4\alpha Z-2\alpha-3\alpha^2+E\right)r
+2\alpha\left(12\alpha Z-2\alpha-9\alpha^2+3E \right)r^2
+4\alpha^2\left(\alpha^2+E\right)r^3
\right]f(r)\\
&+\left[ 6+12\alpha r+4\alpha^2 r^2-8\alpha^3r^3 \right]f'(r)+
r\left[ 3+6\alpha r+4\alpha^2r^2 \right]f''(r)=0.
\end{split}
\end{align}
Equation (\ref{dif1s1s}) is a  second-order 
linear differential equation for $f(r)$. 
To the best of our knowledge, its solution cannot be expressed 
as a combination of the known elementary and/or special functions. 
Since  $r=0$ is a regular singular point\cite{arfken}, at least one solution 
can be found by using the following substitution
\begin{equation}\label{exp0}
f(r)=\sum_{k=0}^\infty c_k r^{k+\rho}.
\end{equation}
Inserting Eq. (\ref{exp0}) into the differential equation, 
collecting terms with the same power of $r$, and
requiring the corresponding coefficients to vanish identically, 
one obtains the indicial equation:
\begin{equation}
3\rho(\rho+1)c_0=0,
\end{equation}
that is used to determine the value of $\rho$. 
Since $f(r)$ must be finite at $r=0$,  we reject $\rho=-1$ and  pick up
$\rho=0$. Setting $\rho=0$ one obtains the first three coefficients:
\begin{align}
\label{c123}
\begin{split}
&c_1=\frac{1}{2}c_0, \\
&c_2=\frac{1}{12} \left(6 \alpha^2-8 \alpha Z-2 E+1\right)c_0, \\
&c_3=\frac{1}{144}\left(32 \alpha^2-32 \alpha Z-8 E+1\right)c_0, \\
\end{split}
\end{align}
and the recursion relation for the remaining ones
\begin{align}
\begin{split}
&\frac{4}{3}c_n \alpha^2(E+1)+
\alpha c_{n+1}\Big[-\frac{4}{3}\alpha-\alpha^2\left(\frac{26}{3}+\frac{8}{3}n\right)+2E+8\alpha Z\Big]\\
&+c_{n+2}\Big[-2\alpha+\frac{1}{3}\alpha^2(2n+1)(2n+7)+E+4\alpha Z\Big]\\
&+c_{n+3}\Big[-1+2\alpha(n+3)(n+4)\Big]
+c_{n+4}(n+4)(n+5)=0.
\end{split}
\end{align}
The value of $c_0$ is arbitrary and can be fixed by imposing 
a normalization condition for the wave function. For the
sake of convenience we put $c_0=1$. 
The first equality in the system (\ref{c123}) is the
cusp condition. It turns out
that the correlation factor obtained from the differential equation (\ref{dif1s1s}) automatically satisfies the
electronic cusp, independently of the values of $\alpha$ and $Z$, 
so that for small $r$ the correlation factor behaves
as $f(r)\sim 1+\frac{1}{2}r$. This result is not surprising. 
The wave function $\Psi$ depends on $r$ through $f(r)$ only, 
so that the factor $f(r)$ alone is responsible for
the cancellation of the $1/r$ singularity 
between the potential and kinetic energy terms.

To obtain the asymptotic form of the solution of the differential equation (\ref{dif1s1s})  we 
keep only the terms proportional to the highest (the third) power of $r$. The resulting equation
\begin{align}
\label{dif1s1s3}
4\alpha^2 f''(r)-8\alpha^3 f'(r)+4\alpha^2\left(\alpha^2+E\right) f(r)=0,
\end{align}
has two linearly independent solutions $e^{(\alpha -\sqrt{-E})r}$ and  $e^{(\alpha +\sqrt{-E})r}$ . The  
acceptable solution is the one with the exponent equal to $\alpha-\sqrt{-E}$ . This suggests the following substitution
\begin{align}
\label{subk}
f(r)=e^{B r} g(r),
\end{align}
where $B$=$\alpha-\sqrt{-E}$.  The differential equation for $g(r)$, obtained  from  Eqs. (\ref{dif1s1s}) and
(\ref{subk}), is:
\begin{align}
\label{diffg}
\begin{split}
&g''(r)\Big[3r+6\alpha r^2+4\alpha^2r^3
\Big]+
g'(r)\Big[6+\left(18\alpha-6\sqrt{-E}\right)r+\left(16\alpha^2-12\alpha\sqrt{-E}\right)r^2\\
&-8\alpha^2\sqrt{-E}\,r^3\Big]+
g(r)\Big[-3+6\alpha-6\sqrt{-E}+\left(6\alpha^2-18\alpha\sqrt{-E}+12\alpha Z-6\alpha\right)r\\
&+\left(24\alpha^2 Z-8\alpha^3-16\alpha^2 \sqrt{-E}-4 \alpha^2\right)r^2
\Big]=0.
\end{split}
\end{align}

We shall present a general method of deriving the first term in the asymptotic expansion of $f(r)$ by using the 
information about the asymptotic behavior of the confluent hypergeometric functions.
When the differential equation is given explicitly, as in the present section, and we know the leading term of
the asymptotic expansion of $f(r)$, it becomes easy to derive the complete asymptotic series. 
Method based on the hypergeometric
functions is even more useful in further sections, where the complete form of the corresponding differential
equation cannot be simply obtained  we confine ourselves merely
to the derivation of the leading term in the asymptotic expansion. For mathematical details of the asymptotic
expansion around an irregular singular point and the dominant balance method we refer to the book of Bender and
Orszag.\cite{bender}

We start by neglecting in Eq. (\ref{diffg}) the terms proportional to $r^0$ and $r^1$. After simple rearrangements one
arrives at the following differential equation:
\begin{align}
\label{diffh}
\begin{split}
&(2\alpha  r +3) h''(r) +2 [4\alpha -(2\alpha  r +3)\sqrt{-E}\, ]h'(r)
- 2 \alpha\, (1+2\alpha+4\sqrt{-E}-6Z) h(r)=0,
\end{split}
\end{align}
The next step is a simple linear change of variables $s= \sqrt{-E} \left(3+2\alpha r\right)/\alpha  $. The
differential equation in the new variable $s$ reads:
\begin{align}
\label{diffkummer}
s\, h''(s)+(4-s)h'(s)+\rho h(s)=0,
\end{align}\\[-6ex]
where\\[-6ex]
 \begin{align}
\label{asym1srho}
\rho=-\frac{1+2\alpha-6Z+4\sqrt{-E}}{2\sqrt{-E}}.
\end{align}
 Equation (\ref{diffkummer})  is a special case of the confluent hypergeometric equation and has two linearly independent solutions expressed usually 
 in terms of Kummer's function\cite{stegun72} $M(-\rho,4,s)$ [denoted also
by $_1F_1$] and Tricomi's function\cite{stegun72} $U(-\rho,4,s)$. The leading terms of the
large-$s$  ($s>0$)  asymptotic expansions of these functions are:\cite{stegun72}
\begin{align}
\label{asymm}
M(a,b,s)&=\frac{\Gamma(b)}{\Gamma(a)}\,e^s
s^{a-b}\left[1+\mathcal{O}\left(\frac{1}{s}\right)\right],\\[1ex]
\label{asymu}
U(a,b,s)&=s^{-a}\left[1+\mathcal{O}\left(\frac{1}{s}\right)\right].
\end{align}
We pick up the normalizable
solution $U(-\rho,4,s)$ and by returning  to the initial variable $r$:
\begin{align}
U(-\rho,4,s)=\left[\frac{\sqrt{-E}}{\alpha}\left(3+2\alpha
r\right)\right]^\rho\left[1+\mathcal{O}\left(\frac{1}{r}\right)\right]\sim
r^\rho\left[1+\mathcal{O}\left(\frac{1}{r}\right)\right],
\end{align}
where the multiplicative constant was neglected since it is irrelevant in the present context. By combining this result
with Eq. (\ref{subk}) one finds that for large $r$
\begin{align}
f(r)= r^\rho e^{(\alpha-\sqrt{-E})r} \left[1+\mathcal{O}\left(\frac{1}{r}\right)\right].
\end{align}
Once the leading term of the asymptotic expansion is known it becomes quite straightforward to obtain the complete
asymptotic series. By inserting the following Ansatz:
\begin{align}
\label{asymfull}
f(r)=r^\rho\, e^{(\alpha-\sqrt{-E})r} \, \sum_{k=0}^\infty \frac{d_k}{r^k},
\end{align}
into the differential equation (\ref{dif1s1s}) and collecting the same powers of $r^{-1}$ one finds that the
indicial equation is automatically satisfied by the choice of $\rho$ given by Eq. (\ref{asym1srho}).  
The recurrence relation determining the $d_k$ coefficients is given by
\begin{align}%
\begin{split} \label{dnrec}
  &d_n [3n(n-1)-(6n-3)\rho+3\rho^2 ]+
d_{n+1}[ -3-12n\alpha\rho+6\alpha(n^2 +\rho^2)+6(n-\rho)\sqrt{-E}\,\, ]\\
  &+ d_{n+2} [6(2\alpha Z-1) +2\alpha^2(2n^2+2n-1) 
 -\alpha^2(8n+4)\rho  
 +4\alpha^2\rho^2 +6\alpha (2n-1)\sqrt{-E}\,\, ]\\ 
  &+d_{n+3} [ 4\alpha^2 (6Z-2\alpha -1) +8\alpha^2(n+1-\rho)\sqrt{-E}\,\,   ]=0,
\end{split}
\end{align}
with $d_0$  arbitrary.  Equation (\ref{dnrec}) is also valid  for $n$=$-$1 and $n$=$-$2 provided that we assume that $d_n$=0  for $n<0$. 
The asymptotic series for the second (unphysical) solution of Eq.~(\ref{dif1s1s}), behaving 
at large $r$ as  $r^{-\rho-4}\,e^{(\alpha+\sqrt{-E})r}\, [1 +{\cal O}(1/r)]$, can be obtained in the same way.  

Summarizing, we found that the correlation factor in Eq. (\ref{psi1}) possesses large-$r$ asymptotic
expansion given by   Eq. (\ref{asymfull})
 with all parameters known analytically as functions of $\alpha$, $Z$,  and $\sqrt{-E}$.
To determine numerical values of $B$ and $\rho$  we performed variational calculations on the series of  helium-like ions using the trial wave function of the form  of Eq. (\ref{psi1}),  with  $f(r)$  represented as  a  15th order polynomial in $r$.  
In this way we obtained  sufficiently accurate values of  $E$ and, consequently,  
of $B = \alpha -\sqrt{-E} $ and of $\rho$  [employing Eq. (\ref{asym1srho})].  
For the value of 
the screening parameter $\alpha$ we adopted: (i) an optimal value for the wave function  of Eq. ~(\ref{psi1}), 
or (ii) the value $\alpha=Z$ corresponding to the  solution for the ``bare-nucleus'' Hamiltonian. Table \ref{table1}
summarizes the
results. We see that, independently of the choice of $\alpha$,   the parameters $B$ and $\rho$  are positive, albeit small. Therefore, somewhat surprisingly, the correlation factor  at 
large $r$  neither decreases to zero as  predicted by Bohm and Pines\cite{bohm53} for the homogeneous electron gas,  nor tends to a constant value as in the standard versions of   F12 
theory\cite{kong12,tenno12}. In fact, it tends to infinity even faster than the linear correlation factor
of the R12  theory of Kutzelnigg and Klopper\cite{kutz85,klopper87}. 

It has to be mentioned that throughout the paper $E$ is treated essentially as a constant. However, $E$ is a
functional of $f$ evaluated with the optimal form of $f$, and thus a function $\alpha$. Nonetheless, this
dependence is rather weak when one is limited to the reasonable vicinity of the optimal value of $\alpha$.

The differential equation (\ref{dif1s1s}) also gives an opportunity to obtain the
correlation
factor with a controlled accuracy for an arbitrary value of $r$. It is clear that the expansion of $f(r)$ in
the powers
of $r$ and the variational minimization gives an access to the short-range part of $f(r)$ but cannot describe
its long-range part with a satisfactory accuracy. On the other hand, the numerical propagation of the differential
equation (\ref{dif1s1s}) can be performed very accurately up to very large distances $r$.  Also the energy  $E$  can be  determined very accurately in this way by adjusting it such that the solution diverging as  $r^{-\rho-4}\,e^{(\alpha+\sqrt{-E})}$  does not show up at large $r$.
We used a high-order Runge-Kutta propagation with a variable step size and  checked carefully the 
convergence of the solution. 
 Figure 1 shows the result of the propagation of the differential equation (\ref{dif1s1s}) for the helium
atom ($\alpha=1.84833$). This numerical propagation result is compared with the variational solution 
expanded in   powers of $r$ up to $r^{15}$. The agreement is very good up to about $r$=8 (the curves in   Fig.~1 are
indistinguishable at $r<6$). At larger distances 
the variational solution becomes completely unrealistic and becomes negative at $r > 14$. 

At large $r$ the propagation curve agrees very well with  the first term  
   of Eq.~(\ref{asymfull}).  
It is remarkable that the leading term of  this  asymptotic expansion  gives  reasonable  approximation to $f(r)$ even for $r$ as small 0.5, 
where the remaining error is slightly less than 7\%. We also found that adding two more terms
from expansion (\ref{asymfull})  significantly 
improves the  approximation around $r$=1, reducing the error   from about 4\% to less
than 0.8\%. Moreover, the reliability of this three-term asymptotic expansion 
  extends  to   $r$=0.2, where the remaining error is  about
5\% (the approximation by the leading term only gives 15\% \ error at this distance).
 These results confirm the validity of the differential
equation (\ref{dif1s1s})  as well as  of the asymptotic form of $f(r)$ given by Eq. (\ref{asymfull}).


\subsection{\label{subsec:1s2s}Correlated Slater orbitals. Triplet state.}

In this  subsection we consider a slightly more complicated model, namely, the  simplest wave function for the  
lowest triplet state of a helium-like ion:
\begin{align}
\label{psi2}
\Psi(r_1,r_2,r)=\left(e^{-\alpha r_1-\beta r_2}-e^{-\beta r_1-\alpha r_2}\right)f(r).
\end{align}
 
The implicit differential equation for $f(r)$ takes the form analogous to Eq. (\ref{symimpl}):
\begin{align}
\label{unsymimpl}
\begin{split}
&\int_0^\infty\!\! \int_{|r_1-r |}^{r_1+r } \, r_1 r_2\,  
 e^{-\alpha r_1-\beta r_2} 
  (\hat{H}-E )  (e^{-\alpha r_1-\beta r_2}-e^{-\beta r_1-\alpha r_2} ) f(r )\, dr_2  dr_1 =0.
\end{split}
\end{align}
The  explicit form of this equation,  obtained easily using the  integral formulas of Appendix A,  
splits naturally into three components  proportional to the exponential factors  
$e^{-2\alpha r}$,   $e^{-2\beta r}$, and  $e^{-2(\alpha  + \beta r)r}$,  respectively.
 Since the differential equation (\ref{unsymimpl}) is symmetric with respect to the exchange $\alpha\leftrightarrow\beta$ we can assume that $\alpha<\beta$.  With this assumption the 
component proportional to the factor  $e^{-2\alpha r}$ dominates at large $r$.
 Neglecting the  two (exponentially) small components one obtains the following equation for $f(r)$:
\begin{align}
\begin{split}
\label{1s2slead}
&f(r)
 \lbrace
4\alpha\beta+ 2 
r\, [ \,(\beta - \alpha Z -\beta Z)(\alpha^2-\beta^2)- \alpha\beta(\alpha^2+\beta^2)-2\alpha\beta E
\,]\\
&+r^2 (\beta^2-\alpha^2) [
 \alpha(3\beta^2-\alpha^2)+2\alpha E +
2Z(\alpha^2-\beta^2) 
 ] \rbrace\\
&+f'(r)4\alpha [-\beta+ (\alpha^2+\beta^2)r+ \beta(\beta^2-\alpha^2)r^2 ]+
f''(r) 2\alpha[-2\beta r-(\beta^2-\alpha^2)r^2  \,]=0.
\end{split}
\end{align}
Neglecting for the moment  terms  proportional to $r^0$ and $r^1$ we obtain  the equation
\begin{align}
\begin{split}
 f(r) [
-\alpha (3\beta^2-\alpha^2)-2\alpha E +
2Z(\beta^2-\alpha^2) 
 ] 
 +4\alpha\beta f'(r)
-2\alpha f''(r)=0,
\end{split}
\end{align}
which has two linearly independent solutions in the form $e^{Br}$ but the only
physically acceptable solution is the one with the exponent $B=\alpha -\gamma$, where  
\begin{align}
\label{bianka2}
\gamma =\sqrt{(\alpha^2-\beta^2)\left(\frac{2Z-\beta}{2\beta}\right)-E}.
\end{align}
Knowing the value of $B$  we can follow the hypergeometric function approach presented in Subsection \ref{subsec:1s1s} and find that the 
  leading term  of the asymptotic expansion for $f(r)$ is  $r^\rho e^{Br}$ with  $B=\alpha -\gamma$ and  
\begin{align}
\label{rhotriplet}
\rho= \frac{\alpha+\beta}{2\beta \gamma} Z 
-\frac{1}{2\gamma}-1 .
 \end{align}
Now keeping all terms in Eq. (\ref{1s2slead}) and  using the  Ansatz (\ref{asymfull})    one obtains the following recursion relation determining the complete 
asymptotic expansion for $f(r)$:
\begin{align}
\begin{split}
&d_{n+3}\left(\alpha^2-\beta^2\right)\Big[ 2Z\left(\alpha^2-\beta^2\right) -2\beta\ \left(E+B^2-2\alpha B\right)+
\beta^2\ \left(\beta^2-3\alpha^2\right) \Big]\\
&+d_{n+2}\Big[
2\left(\alpha^2-\beta^2\right)\left(\alpha-\beta\right)\left( 2B\rho+Z  -4n\beta B - 4\beta B\right)-4\alpha\beta\left(E+B^2-2\alpha B\right)
\Big]\\
&+d_{n+1}\Big[
 8\alpha\beta\rho(\alpha-B)+(4nB+2B -2\beta\rho)  \rho\left(\alpha^2-\beta^2\right)+
4(2n+1)\alpha\beta B\\
&-2n(n+1)\beta\left(\alpha^2-\beta^2\right)-8(n+1)\alpha^2\beta+4\alpha\beta
\Big]
-4\alpha\beta(n-\rho)^2 d_n=0,
\end{split}
\end{align} \\[-3.5ex]
where the value of $d_0$ is arbitrary.

To  confirm  the validity of  formulas derived in this subsection we performed variational calculations using 
 the wave function of Eq.~(\ref{psi2}) and $f(r)$ expanded in powers of $r$ up to $r^{15}$. We used the optimized parameters
$\alpha$=0.321454 and $\beta$=1.968451 which give the energy of $2^3S$ state   $-2.170 104$. This value compares
reasonably with the exact energy of this state equal to  $-2.175 229$. 
With the adopted values of  $\alpha$ and $\beta$, the values of
$B$ and $\rho$, calculated according to Eqs. (\ref{bianka2}) and (\ref{rhotriplet}) are $-0.151753$ and $0.40172$,
respectively. Therefore, in the case of the triplet state $2^3S$, the correlation factor in the wave function
(\ref{psi2}) vanishes exponentially at large distances $r$. This can be understood by invoking the argument that in
the $2^3S$ state the electrons occupy two different shells, so that correlation between them is asymptotically weaker.
Moreover, the Fermi part of the correlation is already included in the zero-order wave function. In
Fig. \ref{figt} we present a comparison of the correlation factors obtained from the numerical propagation and
variational calculation  with the leading term of the asymptotic expansion. The agreement
between the variational result and the numerical propagation is not as good as in Subsection \ref{subsec:1s1s}. This is
due to the slow convergence of the variational result when increasing the number of powers of $r$ included in
the expansion of $f(r)$. Indeed, even with   the $15$th power included, the ratio of first two coefficients in the
expansion of $f(r)$ is equal to $0.367$, while it should be 0.25 (the cusp condition for triplet states). We
were not able to include more powers of $r$ in the variational calculations since the overlap matrix becomes ill
conditioned, and  even in the octuple arithmetic precision the results obtained by symmetric orthogonalization were 
not reliable. The reason for this slow convergence is that for $r=0$ the wave function (\ref{psi2}) vanishes. Therefore,
the energy values are not sensitive to the quality of the trial wave function in the regions close to the
coalescence points of the electrons. Again we find it remarkable that the first term in the asymptotic expansion 
represents $f(r)$ reasonably well in a wide range of distances, although the agreement at intermediate $r$ is not as good as for the singlet  state.

\subsection{\label{subsec:gauss1s}Correlated Gaussian orbital. Singlet state.}

Since the   vast majority of
calculations in quantum chemistry are performed employing  the basis of  
Gaussian orbitals one may ask how the results of previous subsections 
are modified when the orbital basis changes 
from  Slater to  Gaussian functions.
  To investigate this problem we
use the Gaussian analogue of the model from
Subsection \ref{subsec:1s1s}. Namely, we consider the following approximation to the wave function:
\begin{align}
\label{psi4}
\Psi(r_1,r_2,r)=e^{-\alpha r_1^2} \, e^{-\alpha r_2^2 }\, f(r).
\end{align}
It is perfectly clear that the above wave function is a very crude
approximation to the exact one. One can expect, however,  
that this model captures the essential features of   more accurate
approximations  when  the  atomic orbitals are expanded as linear combinations of   Gaussian
functions.  The results obtained for such model extensions
can  easily  be deduced from the equations presented here.

To derive a differential equation for $f(r)$ we start from a suitable  modification 
 (the replacement of $r_1+r_2$ by
 $r^2_1+r^2_2$) of Eq. (\ref{symimpl}).  After changing the variables to $\xi= (r_1+r_2)/r$,  $\eta= (r_1- r_2)/r$ 
and using well-known Gaussian integrals we find that $f(r)$ satisfies the equation
\begin{align}
\label{1s1sgauss}
\left[ 16\,\mbox{Erf}\left(\sqrt{\alpha}r\right)-2+(2E-9\alpha)r+2\alpha^2r^3
\right]f(r)+
\left( 4-4\alpha r^2 \right)f'(r)+2rf''(r)=0,
\end{align}
where $\mbox{Erf}(x)$ is the error function. Since we are   interested
in the  large-$r$  behavior of $f(r)$ we can invoke the asymptotic form of the error function, 
\mbox{Erf}$\left(x\right)=1-e^{-x^2}/(x\sqrt{\pi})+\ldots$,   and replace Eq. (\ref{1s1sgauss}) by
a simpler one 
 \begin{align}
\label{1s1sgaussinf}
\left[ 14+(2E-9\alpha)r+2\alpha^2r^3 \right]f(r)+
\left( 4-4\alpha r^2 \right)f'(r)+2rf''(r)=0.
\end{align}
This equation can be solved exactly in terms  of Kummer and Tricomi functions.   To obtain its   solutions  we   make  the    the substitution
\begin{align}
f(r)=e^{\frac{\alpha}{2}r^2-\gamma r} k(r),
\end{align}
where the parameter  $\gamma$ is yet undetermined. By inserting the above
from of $f(r)$  into Eq. (\ref{1s1sgaussinf})  
  one arrives at the following differential equation for $k(r)$:
\begin{align}
\label{difk}
\left[r \left(-3\alpha+2E+2\gamma^2\right)+14-4\gamma\right]k(r)
+4(1-\gamma r)k'(r)+2rk''(r)=0.
\end{align}
The value of $\gamma$ can be now fixed by requiring that the coefficient
proportional to $rk(r)$  
vanishes identically. Choosing
 \begin{align}
\label{gaussg}
\gamma= \sqrt{\frac{3}{2}\alpha-E} 
\end{align}
  Eq. (\ref{difk}) takes the  form:
\begin{align}
\label{difk2}
rk''(r)+2\left(1-\gamma r\right)k'(r)+\left(7-2\gamma\right)k(r)=0.
\end{align}
Finally, by   change of variable  $x= 2\gamma r$ we transform  Eq.
(\ref{difk2}) into the standard from of the  Kummer   equation\cite{stegun72}
\begin{align}
\label{kummer}
x k''(x)+(2-x)k'(x)-\left( 1 -\frac{7}{2\gamma}\right)k(x)=0
\end{align}
   The two linearly independent solutions  of Eq. (\ref{kummer}) 
are the  Kummer and Tricomi functions,  $M (1- \frac{7}{2\gamma},2,x  )$ and $U  (1-\frac{7}{2\gamma},2,x )$, respectively.  For the same reason as in Section \ref{subsec:1s1s} we pick up the Tricomi  function. Thus, the exact solution of
 Eq. (\ref{1s1sgaussinf})  reads
\begin{align}
f(r)=e^{\frac{\alpha}{2}r^2-\gamma r}\, U\left(1-
\frac{7}{2\gamma},2, 2\gamma r \right).
\end{align}
The asymptotic expansion of the Tricomi function is well-known   [cf. Eq.
(\ref{asymu})], so  the leading term in the large-$r$ expansion of $f(r)$ is:
\begin{align}
\label{asymgau}
f(r)\sim  
r^{ \frac{7}{2\gamma}-1} \, \,  e^{\frac{\alpha}{2}r^2- \gamma r}.
\end{align}
  Since $\alpha$ is positive, $f(r)$ diverges to infinity  large $r$.

 We performed numerical calculations for the helium atom to verify our findings.
 We found variationally that the
optimized parameter $\alpha$ for the wave function (\ref{psi4}) is equal to
$0.859802$. The corresponding energy value
is $E=-2.339039...$  The values of the parameters in Eq. (\ref{asymgau})
that
define the asymptotic expansion are:
\begin{gather}
\gamma= 1.90493,\\
 \frac{7}{2\gamma}-1=0.837342.
\end{gather}
Figure \ref{figgauss} shows the result of the propagation of the
differential equation (\ref{1s1sgauss}) compared with
the leading
term of the asymptotic expansion of $f(r)$. We see a very good agreement
between these two curves at large
interelectronic distances. For comparison, we also plot the correlation
factor obtained from variational
calculations  when $f(r)$  is expanded in powers of $r$. We conclude that the
numerical results presented in Figure
\ref{figgauss} confirm  the analytical results derived in this
subsection.

 
\subsection{\label{subsec:hff12}Correlated SCF  orbitals. Singlet state}

We now consider a more complicated model wave function -- an SCF determinant multiplied by the correlation factor $f(r)$. For simplicity, we will consider only the ground state of the helium like ions. However, the method developed 
here 
can be extended  with   minor modifications to  other states  state of a two-electron atomic system. 
We found it  too tedious to derive   recurrence relations for the  coefficients appearing in the asymptotic
expansion for  $f(r)$. However, we obtained  a relatively 
compact expression for the  first term in this expansion and developed a method to obtain in principle  
as many other terms  as desired. The results of this subsection can be expressed using 
the following theorem 
\begin{theo}
If the wave function for a helium-like ion with charge $Z$ has the form 
 \begin{align}
\label{hff12}
\Psi(r_1,r_2,r)=\phi(r_1)\phi(r_2)f(r),
\end{align}\\[-6ex]
where \\[-6ex]
\begin{align}
\label{hfs}
\phi(r)=e^{-\alpha r}\sum_{k=0}^N c_k r^k,
\end{align}
then the optimal  correlation factor $f(r)$   behaves  at  large  $r$ as  $r^\rho e^{B r}$, with
\begin{gather}
\label{bianka}
B=\alpha-\sqrt{-E}
\end{gather}\\[-6ex]
and\\[-5ex]
\begin{gather}
 \label{rhohf}
\rho=\frac{2N(4Z-\alpha -1) +6Z-2\alpha -1}{2(2N+1)\sqrt{-E}} -2N -2,
\end{gather}
where  $E$ is the variational energy obtained with the wave function $\Psi(r_1,r_2,r)$.
\end{theo}
Note that  we do  not assume  here that the coefficients  $c_k$ are obtained from the solution of 
the matrix SCF equations.   The theorem applies to an arbitrary product
of one-electron functions of the form  of (\ref{hfs}). In fact, the coefficients $c_k$ do not even appear explicitly in
the  equations for the parameters $B$ and $\rho$.

We begin the proof by writing down the analogue of Eq. (\ref{symimpl}). It reads:
\begin{align}
\begin{split}
\label{hf1}
\sum_{k,l,m,n=0}^N c_k c_l c_m c_n
 \int_0^\infty \!\!\int_{|r_1-r  |}^{r_1+r } r^{k+1}_1 r^{l+1}_2  e^{-\alpha(r_1+r_2)}
\left(\hat{H}-E\right) r_1^nr_2^me^{-\alpha(r_1+r_2)} f(r ) dr_2 dr_1  =0.
\end{split}
\end{align}
Similarly as in the  derivations  in Secs. (\ref{subsec:1s1s}) and   (\ref{subsec:1s2s}) we shall  identify the  coefficients that multiply the two highest powers of $r$  in the  differential equation defining  $f(r)$. 
Using  Eq. (\ref{hamilts}) and   Eq. (\ref{asymi2}) we find  that these two highest powers of $r $ 
are $r^{4N+3}$ and $r^{4N+2}$.   This kind of terms can be
produced only by five components of the sum in Eq.~(\ref{hf1}). The component $k$$=$$l$$=$$m$$=$$n$$=$$N$ produces terms of
the order $4N+3$ and $4N+2$, while the   four components for which  $k+l+m+n=N-1$ 
produce terms of the order $4N+2$. 
 As a result, we   need to analyze only  the following two integrals
\begin{align}
\label{matrix1}
M_1=\int_0^\infty \!\!\int_{|r_1-r |}^{r_1+r } r^{N+1}_1 r^{N+1}_2  e^{-\alpha(r_1+r_2)}
\left(\hat{H}-E\right) r_1^Nr_2^Ne^{-\alpha(r_1+r_2)} f(r ) dr_2 dr_1 ,
\end{align}
 \begin{align}
\label{matrix2}
M_2=\int_0^\infty \!\!\int_{|r_1-r |}^{r_1+r } r^{N }_1 r^{N+1}_2  e^{-\alpha(r_1+r_2)}
\left(\hat{H}-E\right) r_1^Nr_2^Ne^{-\alpha(r_1+r_2)} f(r ) dr_2 dr_1 ,
\end{align} 
 which correspond to the $k$$=$$l$$=$$m$$=$$n$$=$$N$ and   $l$=$m$=$n$=$N$,\;$k$=$N$$-$1 case, respectively. 
The remaining three combinations  of indexes  lead to 
the same matrix element as the one given above due to the indistinguishability  
 of electrons and the hermiticity of the Hamiltonian. 
 
 The integrals (\ref{matrix1}) and  (\ref{matrix2}) can be  expressed through the  
integrals $I_{mn}(2\alpha,2\alpha)\equiv I_{mn}$  of Appendix A.  Making use of  the asymptotic relation (\ref{asymi2}) one easily finds that  
\begin{align}
\label{m14n4}
\begin{split}
M_1=&-  f''(r)I_{2N+1,2N+1}     -  r^{-1} f'(r)
\left[-\alpha I_{2N+2,2N+1}-\alpha r^2 I_{2N,2N+1}+\alpha I_{2N,2N+3}\right]\\
-&    f(r)(\alpha^2 +E)I_{2N+1,2N+1}  + \mathcal{R}_{4N+2},
\end{split}
\end{align}
where $\mathcal{R}_{4N+2}$ collects terms involving $r^{4N+2}$ and lower powers of $r$.
More explicitly, 
\begin{align}
\label{m14n4c}
\begin{split}
M_1=&-\frac{e^{-2\alpha r}}{2\alpha}r^{4N+3}
\big[f''(r)C_{2N+1,2N+1}+ \alpha
  f'(r) \big(
  C_{2N,2N+3}-  C_{2N+2,2N+1}  -  C_{2N,2N+1}\big)\\
&+(\alpha^2 +E)C_{2N+1,2N+1}f(r) +\mathcal{O} (r^{-1} ),
\big] ,
\end{split}
\end{align}
where $C_{nm}$ are the coefficients  appearing in Eq. (\ref{asymi2}) and given by Eq.  (\ref{cmn2}). 
Noting that  
\begin{equation}\label{lem1}
C_{2N+2,2N+1}+C_{2N,2N+1}-C_{2N,2N+3}=2C_{2N+1,2N+1}
\end{equation}
end  equating  the coefficient  at $r^{4N+3}$ to zero we obtain the equation 
\begin{align}
\label{lead2}
0=f''(r)-2\alpha f'(r)+    f(r)(\alpha^2+E).
\end{align}
which is a strict analogue of Eq. (\ref{dif1s1s3}).  Its solutions are $e^{(\alpha + \sqrt{-E})r}$
and  $e^{(\alpha-\sqrt{-E})r}$, the latter one being  the only acceptable choice. 

To obtain the preexponential factor we follow the method used in   in Section \ref{subsec:1s1s} and make 
the substitution $f(r)=e^{Br}g(r)$, where $B=\alpha-\sqrt{-E}$. To derive a useful equation 
for $g(r)$ we need a more accurate representation of the the l.h.s. of Eq.(\ref{hf1}) than  that given by Eq. (\ref{m14n4c}). 
The required equation,   including the next lower power of $r$,   has been derived in Appendix B. It has the form
\begin{align}
\label{hfdif1} 
\begin{split}
&  r \big[f(r)(\alpha^2+E)-2\alpha f'(r)+f''(r)\big](2N+1)\\
+&f(r )\Big\lbrace  (4N+3)\big[2Z 
- \frac{\alpha}{2}(2N+3)  
+ \frac{E}{2\alpha}(2N+1) 
 +4b_{N}(\alpha^2+E)\big] - \Big\rbrace \\
+&f'(r )\Big[ (2N+1)-8\alpha b_{N}(4N+3) \Big]+f''(r)(4N+3)\big[\frac{2N+1}{2\alpha} +4b_N\Big]=0,
\end{split}
\end{align}
where $b_N=c_{N-1}/c_N$. After the substitution $f(r)=e^{Br}g(r)$ we obtain  the following differential equation for $g(r)$:
\begin{align}\label{gEq}
\begin{split}
-&2\alpha   \Big[1+4\sqrt{-E}+2\alpha(N+1)+4N(2N+3) \sqrt{-E}-2Z(4N+3) \Big]g(r)\\
-&2\Big[ (4N+3)\sqrt{-E} (2N  +1 +8\alpha b_{N }) -2\alpha   (2N+1) ( 2N +2-\sqrt{-E}\,r ) 
 \Big] g'(r)\\
+&\Big[ 8\alpha b_{N}(4N+3)+ (2N+1)( 4N+3+2\alpha r) \Big]g''(r)=0.
\end{split}
\end{align}
If we now introduce a new variable $x=2\sqrt{-E}(r+a)$, where
 \begin{align}
a=\frac{ (4N+3)\left(2N+1 +8\alpha b_N \right)}{2\alpha (2N+1) }
\end{align}
then Eq.  (\ref{gEq}) reduces to the  standard   Kummer's differential equation
 \begin{align}\label{Kumm2}
\begin{split}
x g''(x)+ (4N+4-x) g'(x)+ \rho g(x)=0,
\end{split}
\end{align}
with  $\rho$  given  now by Eq. (\ref{rhohf}). Note that when $N=0$, Eq. (\ref{Kumm2}) reduces to 
Eq. (\ref{diffkummer}) with  $\rho$  given by  Eq. (\ref{asym1srho}).  Using the asymptotic representation 
of the Tricomi function, Eq.  (\ref{asymu}),   we find that  $g(r) \sim r^{\rho}$   and 
$f(r)\sim r^{\rho} e^{Br}$ at large  $r$,  where $B$ and $\rho$ are given by Eqs.   (\ref{bianka})  and (\ref{rhohf}). 
 The complete large-$r$ asymptotic expansion of $f(r)$ can be obtained by inserting  the Ansatz 
of Eq.  (\ref{asymfull}), with $B$ and $\rho$ given by Eqs. (\ref{bianka})  and (\ref{rhohf}),   into the differential equation for $f(r)$ and deriving recurrence relation for the coefficients $d_n$. 
Because of its great complexity we did not attempt to carry out this procedure except for
 $N=1$ and $N=2$.  This completes the proof of the Theorem formulated at the beginning of this section.
 

We find it remarkable  that the value of $B$ does not depend explicitly
on $N$. One might expect that an increase of $N$ changes the orbital part of the wave function 
significantly at large $r$ and, in turn, changes the rate of the asymptotic growth of $f(r)$.
This intuition seems to be invalid and $B$ is found to be a universal parameter, dependent 
on the orbital part of the wave function through the values of $\alpha$ and $E$ only.
There is of course an implicit dependence  on $N$ through the value of $E$. 
This dependence is found to be very weak since the
energy saturates very quickly with increasing $N$.
For example,   for  the helium atom with the optimized parameter
$\alpha=1.84833$  our best theoretical value of $B$, based on the energy extrapolation toward the complete
basis (\emph{i.e.} infinite $N$) is $0.148505$, while the values obtained with $N=2,3,4$ are $0.148463$, $0.148521$, and $0.148504$, respectively. 
Even the value corresponding to $N=0$ ($0.147961$) compares well with the estimated limit.
Similar conclusions can be drawn from the calculations on the helium-like ions. 
Therefore, the parameter $B$ seems to be universal and weakly dependent on the quality of the ``orbital'' part of the wave function.

The dependence of $\rho$ on $N$ appears to be rather strong.  At large $N$  this parameter decreases linearly 
with $N$  with the slope of $-2$:
\begin{align}
\label{skewrho}
\rho =    -2N -2 +\frac{4Z - 1 -\alpha}{2\sqrt{-E} }  +\mathcal{O} \left({1\over N} \right) ,
\end{align}
 This result is independent of the values of $E$, $\alpha$ and $Z$.  
 Figure \ref{figrho} presents the shape of $\rho(N)$ calculated for
the helium atom with an optimized parameter $\alpha$. 
One can see that the convergence toward the linear 
asymptote is fast, so that even for $N$ being as small as $3.0$ the error resulting from the use of Eq.
(\ref{skewrho}) is of the order of $1\%$. Therefore, for 
longer expansions of $\phi(r)$, the approximation (\ref{skewrho}) is sufficiently accurate for all practical purposes. 

To verify our findings numerically, we derived explicit differential equation for $f(r)$ in the case of $N=2$, i.e.,  a three-term SCF orbital  used with  in Eq. (\ref{hfs}).
With the  optimized parameter $\alpha=1.920904$ and $N=2$ we obtained the SCF energy equal to
$-2.86159 $ which compares well with the  Hartree-Fock limit\cite{szalewicz81} of   $-2.86168$.
 Figure \ref{fighf} presents results of
the numerical propagation of the differential equation for $f(r)$ in the described case. For comparison, we plot
the results of the variational calculations with $f(r)$ expanded in a basis set of the powers of $r$. Excellent
agreement between those curves is found for small $r$ albeit for a medium range the variational result
becomes unstable and progressively less accurate. A new feature of the correlation factor in the present example is that
it is no longer monotonic over the whole domain, as found in the previous models. Instead, it possesses a single maximum
for a small $r$ value and then a shallow minimum somewhere at the medium large. The leading term of the asymptotic
expansion of $f(r)$ is $r^\rho e^{Br}$ with $B=0.220361$ and $\rho=-4.38436$, calculated according to
Eqs. (\ref{bianka}) and (\ref{rhohf}). Satisfactory agreement between this term and the propagation curve is found for
larger values of $r$.


\subsection{\label{subsec:kutzf12}The Kutzelnigg Ansatz}

  In this subsection we extend our  approach by considering the following   Ansatz:\vspace{-5ex}

\begin{align}\label{KutzAnsatz}
\Psi(r_1,r_2,r)=\Psi_0(r_1,r_2)f(r)+\chi(r_1,r_2,r),
\end{align}
\vspace{-5ex}

\noindent
where $\Psi_0(r_1,r_2)$ is a reference function (either a product of simple exponential functions or   SCF 
orbitals) and  the complementary function $\chi(r_1,r_2,r)$ is an ordinary expansion in a  set of orbital products. This form of the
wave function  with $f(r)$ chosen as $1+\frac{1}{2}r$ was used by Kutzelnigg in his work on the R12 theory\cite{kutz85}.
 To simplify derivations we assume that  the complementary wave function
$\chi(r_1,r_2,r)$ is restricted to the following form
\begin{align}\label{compl}
\chi(r_1,r_2,r) =e^{-\alpha(r_1+r_2)}\sum_{kl}^{M} d_{kl}\,r_1^k r_2^l.
\end{align}
The  basis set used  in the expansion (\ref{compl}) is incomplete due to lack of angular functions.  
Including them (via even powers of $r$) is straightforward and we shall show later that it will 
not affect the asymptotic behavior of   $f(r)$.   To avoid technical complications we   
make the choice $\Psi_0(r_1,r_2)=e^{-\alpha(r_1+r_2)}$. 
The main result of this section can be formulated as follows:
\begin{theo}
If the  wave function for the helium like ions has the form 
\begin{align}
\label{kutz1}
\Psi(r_1,r_2,r)=e^{-\alpha(r_1+r_2)}f(r)+e^{-\alpha(r_1+r_2)}\sum_{kl}^{M} d_{kl}\,r_1^k r_2^l,
\end{align}
then  the optimal  correlation factor $f(r)$   behaves  at  large $r$     as  $r^\rho  e^{B r}$, 
where  $\rho$ and $B$   are given by Eqs. (\ref{asym1srho}) and (\ref{bianka}), i.e.,  
are   the same as in the case of the wave function of Eq. (\ref{psi1}).  
\end{theo}
To prove this theorem we have to analyze a differential equation for $f(r)$. Such an equation is obtained  by inserting  Eq. (\ref{KutzAnsatz}) into the Rayleigh-Ritz functional, evaluating its
functional derivative with respect to $f(r)$ and equating this derivative to zero. The resulting
equation reads:
 \begin{align}
\label{inh0}
\begin{split}
&\int_0^\infty \!\!\int_{|r_1-r|}^{r_1+r} \, r_1 r_2\, e^{-\alpha(r_1+r_2)} \left( \hat{H}-E \right)
e^{-\alpha(r_1+r_2)} f(r)dr_2dr_1 =\\
-&\int_0^\infty \!\! \int_{|r_1-r|}^{r_1+r}  r_1 r_2\, e^{-\alpha(r_1+r_2)} \left( \hat{H}-E \right)
\chi(r_1,r_2)dr_2dr_1 .
\end{split}
\end{align}
We assume here that the linear coefficients $d_{kl}$ on the r.h.s. are fixed and have already been optimized by solving appropriate algebraic equations involving the optimal $f(r)$. 

The homogeneous, left-hand side of the above equation is the same as in Eq. (\ref{symimpl}), except for an additional factor of   $-e^{-2\alpha r}/(48\alpha^3)$.  
The inhomogeneity on the r.h.s., which we will further denote by $G(r)$, can  
be easily expressed  through the combinations of auxiliary integrals 
$I_{mn}(2\alpha,2\alpha)\equiv I_{mn}$  evaluated in  Appendix A. The result reads:
\begin{align}
\begin{split}
\label{inh2}
G(r)&=\sum_{kl}^M
d_{kl}\big[\left(\alpha^2+ E  -{1\over r}\right)I_{k+1,l+1} -\alpha(k+1)I_{k,l+1}  
 -\alpha(l+1)I_{k+1,l}\\ 
& +\frac{1}{2}k(k+1)I_{k-1,l+1} + \frac{1}{2}l(l+1)I_{k+1,l-1} 
+Z(I_{k,l+1} +I_{k+1,l})   \big].
\end{split}
\end{align}
According to Eq.~(\ref{asymi2}) from the Appendix A each of the integrals $I_{mn} $
appearing in the equation above is  a finite order polynomial in $r$ multiplied by the exponential function
$e^{-2\alpha r}$. Therefore, the inhomogeneity $G(r)$ is also a    polynomial 
 [of the $(2M +3)$th order] times  $e^{-2\alpha r}$. Substituting this form of $G(r)$ into Eq.(\ref{inh0}), 
using Eq. (\ref{symimpl}) to represent the homogeneous part of Eq.(\ref{inh0})  and canceling the exponential 
factors we find the following differential equation for $f(r)$:
 \begin{align}
\begin{split}
\label{inh4}
&\big[
-3+3\left(4\alpha Z-2\alpha-3\alpha^2+E\right)r
+2\alpha\left(12\alpha Z-2\alpha-9\alpha^2+3E \right)r^2 
 +4\alpha^2\left(\alpha^2+E\right)r^3
\big]f(r)\\&+
\big[ 6+12\alpha r+4\alpha^2 r^2-8\alpha^3r^3 \big]f'(r) 
 +r\big[ 3+6\alpha r+4\alpha^2r^2 \big]f''(r)=
-48\alpha^3\sum_{k=0}^{2M+3} g_k\, r^k.
\end{split}
\end{align}
\vspace{-4ex}

\noindent 
where the   coefficients $g_k$ can 
be easily expressed through $d_{kl}$ and the $C_{mn}$ coefficients of Appendix A.

It is  known  that the general solution of an  inhomogeneous
differential equation is given by a linear combination of the solutions of the  homogeneous
problem plus any particular solution. To find this particular solution,  denoted by $f_S(r)$, we try  a finite 
order polynomial as an educated guess 
\begin{align}
\label{fskutz}
f_S(r)=\sum_{k=0}^{2M+3} h_k\, r^k.
\end{align}
Equations determining the coefficients $h_k$ are found by inserting the above Ansatz into the differential equation (\ref{inh4}) and
gathering the factors multiplying the same powers of $r$. The first three of these equations are 
\begin{align}
\begin{split}
&-h_0+2h_1+16\alpha^3g_0=0,\\
&3h_0\left(-2\alpha-3\alpha^2+E+4\alpha Z\right)+h_1\left(4\alpha-1\right)+6h_2+16\alpha^3g_1=0,\\
&\alpha h_0\left(-4\alpha-18\alpha^2+6E+24\alpha Z\right)+h_1\left(-6\alpha-5\alpha^2+3E+12\alpha Z\right)\\
&+3h_2\left(12\alpha-1\right)+36h_3+48\alpha^3g_2=0
\end{split}
\end{align}
and the general form is 
\begin{align}
\begin{split}
&4\alpha^2\left(\alpha^2+E\right)h_n-2\alpha\left[2\alpha+\left(13+4n\right)\alpha^2-3E-12\alpha Z\right]h_{n+1}\\
&+\left[-6\alpha+\alpha^2(2n+1)(2n+7)+3E+12\alpha Z\right]h_{n+2}+\left[-3+6\alpha(n+4)(n+3)\right]h_{n+3}\\
&+3(n+4)(n+5)h_{n+4}+48\alpha^3g_n=0.
\end{split}
\end{align}
The number of equations is the same as the number of coefficients and  the determinant of the system 
of equations does not vanish.  Having found the   special solution $f_S(r)$,
we can write the general solution of Eq. (\ref{inh4}) 
 \begin{align}
\label{inhsol1}
f(r)=c_1 f_1(r)+c_2 f_2 (r)+f_S(r),
\end{align}
where $f_1(r)$ and $f_2(r)$ are the solutions of the homogeneous problem behaving asymptotically 
as, $e^{(\alpha-\sqrt{-E})r}r^{\rho }$   and $e^{(\alpha+\sqrt{-E})r}r^{\rho'}$, respectively, see the discussion around Eqs.  (\ref{asymm})-(\ref{dnrec}) in  Sec.  \ref{subsec:1s1s}. 

 We can  fix the value of $c_2$ as equal to zero, otherwise the wave function would not be
normalizable. Thus, the long-range behavior of $f(r)$ in the present case reads:
\begin{align}
f(r)  \sim  c_1 e^{(\alpha-\sqrt{-E})r}r^{\rho }+f_S(r),
\end{align}
where $c_1$ can  be fixed by  normalization. 
Since the particular solution is characterized by a
polynomial growth and the chosen solution of the   homogeneous   problem grows exponentially, the leading term
of the asymptotic expansion remains exponential. In other words, for a sufficiently large $r$  the behavior of $f(r)$ is always dominated by the exponential growth of the solution to the  homogeneous problem.
 This formally completes the proof of the theorem stated at the beginning of this Section.

It is easy to extend the above theorem by including higher angular momentum functions in the one-electron
basis set. One can show that this is equivalent to taking the following form of the complementary wave function
\begin{align}
\begin{split}
\chi(r_1,r_2,r)&=
  e^{-\alpha(r_1+r_2)}\sum_{kl} d_{kl}^{(0)}\,r_1^k r_2^l+
r^2 e^{-\alpha(r_1+r_2)}\sum_{kl} d_{kl}^{(1)}\,r_1^k r_2^l\\
&+r^4 e^{-\alpha(r_1+r_2)}\sum_{kl} d_{kl}^{(2)}\,r_1^k r_2^l+\ldots
\end{split}
\end{align}
This extension does not change the main feature of the differential equation that was used in the proof. Namely, the
  solution  of the homogeneous problem remains unchanged 
  and the inhomogeneity is still a finite-order polynomial in $r$. Therefore, a special solution has the
polynomial character and does not contribute to the leading term in the long-range asymptotics. 

We also  considered another   variant of the Kutzelnigg Ansatz:
\begin{align}
\label{kutzab}
\Psi(r_1,r_2,r)=e^{-\alpha(r_1+r_2)}f(r)+e^{-\beta(r_1+r_2)}\sum_{kl}^{M} d_{kl}\,r_1^k r_2^l,
\end{align}
which differs from the wave function (\ref{kutz1}) by the choice of different exponent in the complementary part $\chi(r_1,r_2,r)$  of the wave function. This additional flexibility is not 
very effective in the calculations on the helium atom. We checked that the optimal value of $\beta$ is very close to the adopted value of $\alpha$ and the energy gain is insignificant. However, when passing to many-electron systems and using the
expansion of pair functions similar to Eq. (\ref{kutzab}), the splitting of $\alpha$ and $\beta$ corresponds to
the use of more diffuse (or more tight) basis set functions in $\Psi(r_1,r_2,r)$ than in $\Psi_0(r_1,r_2)$. This is an important case and therefore the model (\ref{kutzab}) is worth considering.
 As before, the extension of (\ref{kutzab})
by including higher angular momentum functions is simple, so we proceed only with $s$-type functions in the basis.

By repeating the derivation in the previous model, Eqs. (\ref{inh0})-(\ref{inh4}),  we find that the  differential equation for $f(r)$ is the same as Eq. (\ref{inh4}), except that the inhomogeneity
in  Eq. (\ref{inh4}) is now given  by the function 
\begin{align}
\begin{split}
\label{inh5}
 \widetilde{G}(r)=
-48\alpha^3\,e^{-2(\beta-\alpha)r}\sum_{t=0}^{2M+3} \widetilde{g}_k\, r^k,
\end{split}
\end{align}
where $\widetilde{g}_k$ are defined in the same way as the $g_k$ coefficients in  Eq. (\ref{inh4}). 
The solution of the homogeneous problem is the
same as in Subsection \ref{subsec:1s1s}. We also found that with appropriate choice
 of $\widetilde{h}_k$  the function 
\begin{align}
\label{fskutzab}
\widetilde{f}_S(r)=e^{-2(\beta-\alpha)r}\sum_{k=0}^{2M+3} \widetilde{h}_k\, r^k,
\end{align}
is a particular solution of the full equation containing the inhomogeneity $\widetilde{G}(r)$. 
We can thus use the same  arguments as    
previously and infer that   
\begin{align}
\label{inh6}
f(r)\sim c_1 e^{(\alpha-\sqrt{-E})r}r^{\rho}+e^{-2(\beta-\alpha)r}\sum_{k=0}^{2M+3} \widetilde{h}_k\, r^k,
\end{align}
asymptotically  for large $r$. 
The dominant  term of this formula depends on the relation between $\alpha$ and $\beta$.
In particular the large-$r$  the asymptotics of $f(r)$  is given by 
\begin{align}
&f(r)\sim r^{\rho}\, e^{(\alpha-\sqrt{-E})r} \hspace{-9em} &\mbox{for}\;\;\;  \beta > \beta_c,\\[1ex]
&f(r)\sim r^{2M+3}\, e^{ 2(\alpha -\beta )r} \hspace{-9em}&\mbox{for}\;\;\; \beta <\beta_c, 
\end{align}
where $\beta_c$ is the critical value of $\beta$ equal to
\begin{align}
\beta_c=\frac{1}{2}(\alpha+\sqrt{-E}).
\end{align}
 Thus, independently of the choice of $\beta$ we find an exponential growth of $f(r)$ at large $r$.

\section{\label{sec:numerical}Discussion and conclusions}

\subsection{\label{subsec:anal}The ``range-separated'' model of the correlation factor}

The analytic results  presented in the previous section
can be put into practical use only if a simple analytical form of the correlation factor can be found 
that mimics, to a good approximation, the exact behavior of $f(r)$  both at small and at large interelectronic distances $r$. This goal is far from being straightforward. 
This is mainly due to considerable  change in the shape of the correlation
factor when the function $\Psi_0$  is modified. For the simplest possible $\Psi_0$ taken as the product of $1s$
orbitals the correlation factor is a monotonically growing function, while  
for $\Psi_0$ taken as  an SCF determinant,  $f(r)$  exhibits a maximum and   minimum
before the onset of the monotonic exponential growth. 
Knowing the behavior  of the correlation factor at small and large $r$  we can propose a 
``range-separated" form  with a Gaussian switching   
\begin{align}
\label{rs1}
f(r)=\big(1+\frac{1}{2} r\big) \,e^{-\mu r^2}+ c\, r^{\rho}\,e^{Br}\, S_n(\mu r^2),
\end{align}
where
\begin{align}
\label{switch}
S_n(x) = 1 -e^{-x } \sum_{l=0}^n \frac{x^{ l}}{l!}
\end{align}
serves as the ``switching function'' that interpolates smoothly between the two
regimes and the switching is controlled  by adjustable parameters $c$  and $\mu$. 
To eliminate the singularity appearing when $\rho <0$  we take as $n$ the smallest integer 
satisfying  $2n +\rho \ge 0$.  For positive $\rho$ we set $n=0$.
This form of $f(r)$ is slightly reminiscent of the error-function based range-separation of the 
Coulomb interaction in the density functional  theory\cite{leininger97}.  
We can increase  somewhat the flexibility of this representation by using the Ten-no's factor 
at short range:
\begin{align}
\label{rs2}
f(r)= \frac{1+2\gamma-e^{-\gamma r}}{2\gamma}\,e^{-\mu r^2}  +  c \, r^{\rho}\,e^{Br}\, S_n(\mu r^2).
\end{align}
We found that, the analytical form (\ref{rs2}) is very flexible. By means of the optimization of the
adjustable parameters we are able to obtain a very good analytic fit for each correlation factor  discussed in the paper.

When the correlation factor of the form (\ref{rs2}) is used in the calculations, new classes of
the two-electron integrals arise that were not considered in the literature so far. In these integrals the factors $r^{\rho}$, $e^{-a r}$ and $e^{-a r^2}$ are present collectively. For the atomic calculations we managed to express these
integrals in terms of the incomplete Gamma and error functions, both in the Slater and Gaussian one-electron basis,
and implement them efficiently. These integrals become substantially more difficult when one passes to the many-center molecular systems. The work on evaluating them   is in progress 
in our laboratory.
 
\subsection{\label{subsec:examp}Results of exemplary calculations  }

To check the effectiveness of the ``range-separated" representation  of Eq.~(\ref{rs1}) 
and Eq.~(\ref{rs2}) we performed variational 
calculations with the wave function of the form of Eq. (\ref{psi1}) and  (\ref{kutz1}).  
The values of the parameters $B$ and $\rho$ were  fixed according to Eqs. (\ref{bianka}) and (\ref{asym1srho}).
The exponent $\alpha$ was set equal to 1.84833.
The parameters $\gamma$, $\mu$ and $c$ were obtained by a least square fit to the exact correlation factor in 
Eq.~(\ref{psi1}),  obtained from the numerical solution  
of  Eq.~(\ref{dif1s1s}).  We found that for the helium atom $\gamma = 0.209587$, $\mu = 0.448695$ and $c=1.170940$ are
optimal when Eq.~(\ref{rs2}) is used, whilst for Eq.~(\ref{rs1}) the values $\mu = 0.861347$ and $c=1.169033$ are
appropriate.

The results are summarized in Table \ref{table2}. 
An inspection of  this table shows that accounting for the correct 
large-$r$ behavior of $f(r)$ via simple formulas of Eq.~(\ref{rs1}) and  Eq.~(\ref{rs2}) improves significantly the
energies   obtained with the standard  R12 or F12  correlation factors.  
As expected, the improvement  is smaller when the exponential factor with optimized $\gamma $ is used. 
Note, however, that the optimal value of $\gamma$, equal to 0.2,  is in this case much smaller than the value recommended in standard F12 calculations\cite{noga08}. 
 
It can be seen  that with the
correlation factor of the form (\ref{rs2}) used in the wave function  of Eq.~(\ref{psi1}) we recover about 70\% of the
correlation energy, so that the expansion in a set of excited state determinants is required only for the remaining 30\%. Standard R12 approximation is  worse in this respect, recovering about 60\% \ of the correlation energy. 

When the wave function of the form of Eq.~(\ref{kutz1}) is used in the calculations,  
the obtained energy differences are much smaller but  one can see that  including the correct asymptotics of $f(r)$
always improves the results.  It should be pointed out  that  in this case the parameters of the correlation factors 
of Eqs.~(\ref{rs1}) and  (\ref{rs2})  were optimized  for the wave function of Eq.~(\ref{psi1}).
Nevertheless, the difference between the energy obtained with the approximate correlation factor of 
Eq.~(\ref{rs2})  and the fully optimal one, equal to  0.18 milihartree, is smaller than the 
corresponding difference remaining when using the wave function of  Eq.~(\ref{psi1}). 
It may also be noted that the energy obtained  with the optimal wave function of Eq.~(\ref{kutz1}), i.e, with orbitals of $s$-type symmetry only,  is slightly better than the energy  from the full CI calculations in the saturated $spdf$ basis set\cite{bunge70,carrol79,weiss61}.  With the linear correlation factor  the $spd$ limit would be  reached with this wave function.  

We also performed calculations with Ten-no's, exponential  correlation factor and  several values of $\gamma$ which are usually recommended in
the literature with $\gamma=1.0$ being the most common choice.\cite{noga08,skomo11}.
Other values, $\gamma=0.5$ and $\gamma=1.5$ were also employed\cite{yousaf08,rauhut09,shiozaki09}. 
The results are shown  in Table \ref{table2}. On can see that all these choices of $\gamma$  give
 results worse than the ``range-separated'' correlation factor of Eq.~(\ref{rs2}).  However, when the exponential correlation factor with optimal $\gamma$ is used in the wave function of  Eq.~(\ref{kutz1}) the energy is slightly    
better than the one obtained with the asymptotically corrected linear correlation factor of Eq. ~(\ref{rs1}).  
This is the manifestation of the superiority of the Ten-no's factor over the linear one at intermediate interelectronic distances.

\subsection{\label{sec:conclusions}Summary and conclusions}
In this work we have considered  the problem of an optimal form of the correlation factor $f(r)$ for explicitly 
correlated wave functions, specifically, its asymptotic behavior at large interelectronic distances $r$. We employed the helium atom and helium-like ions as model systems 
and studied several approximate forms of the the wave function. 
For the simplest
case of the wave function of the form $e^{-\alpha (r_1+r_2)}f(r)$ the optimal correlation factor is exponentially  growing
function with no extremal points at short range. On the other hand, for the case of an SCF  determinant multiplied by the correlation factor, $f(r)$ possesses a single maximum in a small $r$ regime and a minimum at medium
$r $ distances. However, in both cases the asymptotic form of the correlation factor is $  r^\rho e^{Br}$, 
with $B >0$, so that at large interelectronic  distances $f(r)$ diverges
exponentially. While the presence of a maximum in the correlation factor for the SCF case has been observed in
the study of Tew and Klopper\cite{tew05}, neither the presence of the minimum nor the large-$r$ divergence of $f(r)$ have been noticed.  

We  presented a method to derive a well-defined differential equation for $f(r)$ that can be solved analytically in
the large-$r $ regime or alternatively integrated  numerically with arbitrary precision using well-developed 
propagation techniques. 
The exact analytic information about its  solution gives us an opportunity to design new functional  form for the
correlation factor. We proposed a ``range-separated'' model where the short-  and long-range   regimes are
approximated by different formulas and sewed together by using a switching function. Simple exemplary calculations with
the  new form of the correlation factor show that it performs significantly better than the correlation factors used in R12 or F12 methods.

The method proposed in this paper can be a subject to several extensions. First of all, it can be
applied to a two-center system  to reveal the possible dependence of
the correlation factor on the internuclear distance. The second extension goes towards the three-electron atomic
systems, such as the lithium atom. This extension may shed some light on the problem of ``explicit correlation of triples'' considered recently in the literature\cite{kohn09,kohn10}

To apply the proposed form of the correlation factor in calculations for molecular systems,  
difficulties concerning the evaluation of the new integrals and application of the RI approximations must be addressed. The work in this direction is in progress in our laboratory. We hope that the proposed models of $f(r)$ will find applications in explicitly correlated atomic and molecular calculations and will help to increase  the 
accuracy of these calculations. 

\begin{acknowledgments}
This work was supported by the NCN grant  NN204182840. ML thanks  the Polish Ministry of Science and Higher Education for the support through the project ``Diamentowy
Grant'',  number DI2011 012041.   RM thanks the Foundation for Polish Science for support within the MISTRZ
programme. 
Part of this work was
done while RM was visiting  the Kavli Institute for Theoretical Physics, University of California at Santa Barbara
and was supported by the  NSF grant PHY11-25915.
\end{acknowledgments}

\newpage
\appendix
\section{Evaluation of auxiliary integrals}
\label{app:appa}
In this Appendix we  give  expressions for the  integrals:
\begin{align}
\label{a1}
I_{mn}(\alpha,\beta,r)=\int_0^\infty \!\! \int_{|r_1-r|}^{r_1+r} e^{-\alpha r_1-\beta r_2} r_1^m r_2^n dr_2dr_1,
\end{align}
which appear in the derivation of differential equations for $f(r)$. 
We will assume that $m$ and $n$ are non-negative integers and that  $\alpha+\beta>0$.
The closed form expressions for the integrals (\ref{a1}) can be obtained most easily 
by the change of   variables $\xi= (r_1+r_2)/r$,  $\eta= (r_1- r_2)/r$  and the appropriate change of integration range to  $\xi\in[1,+\infty]$ and $\eta\in[-1,+1]$.   The absolute value of the Jacobian is
$|J|=r^2/2$. 
The integral (\ref{a1}) can now be  written  as:
\begin{align}
\label{a7}
I_{mn}(\alpha,\beta,r)=\sum_{l=0}^m\sum_{k=0}^n {m \choose l}{n \choose k}(-1)^{n-k}J_{k+l,m+n -l -k}(\alpha,\beta,r),
\end{align}
where  
\begin{align}
 \label{Jkl}
J_{kl}(\alpha,\beta,r) =
 2\left(\frac{r}{2}\right)^{k+l+2}\, A_k(p )\, B_l(q ),
\end{align}
  $A_k(p)$ and $B_k(q)$  being  the well-known integrals:
\begin{align}
\label{Ak}
A_k(p)  =& \int_1^{\infty} \xi^k e^{-p\xi} d\xi = \frac{k!}{p^{k+1}} e^{-p} \sum_{j=0}^k \frac{p^j}{j!},\\
 \label{Bl}
B_l(q)  =& \int_{-1}^1 \eta^l e^{-q\eta} d\eta =
 \frac{l!}{q^{l+1}}\left[ e^{q} \sum_{j=0}^l \frac{(-1)^jq^j}{j!}-e^{-q} \sum_{j=0}^l \frac{q^j}{j!}\right]
\end{align}
computed at $p$=$ r (\alpha+\beta)/2$  and  $q$=$r(\alpha -\beta)/2$. When $\alpha$ =$\beta$, i.e., 
$q$=$0$  then  
\begin{equation} 
\label{B0}
B_l(0)  = 
 \frac{1}{l+1} \left[ 1 +(-1)^l\right] .
\end{equation}
 
In Sec. \ref{subsec:hff12} we  need information about  the large $r$ behavior of the integrals $I_{mn}(\alpha,\alpha,r)$.  
Using Eq.  (\ref{Jkl}) we find 
\begin{align}
\label{asymj}
J_{kl}(\alpha,\alpha,r)=\frac{e^{-\alpha r}}{\alpha}\left(\frac{r}{2}\right)^{k+l+1}\frac{1+(-1)^l}{1+l}\left [1 + \frac{k}{\alpha r} +
\mathcal{O}\left(\frac{1}{r^2}\right) \right].
\end{align}
Inserting this result  into Eq. (\ref{a7}) and rearranging summation order we arrive at
\begin{align}
\label{asymi2}
I_{mn}(\alpha,\alpha,r)=\frac{e^{-\alpha r }}{\alpha}r^{m+n+1} \left[ C_{mn}+
\frac{D_{mn}}{2\alpha r}  +{\mathcal  O}\left(\frac{1}{r^2}\right) \right],
\end{align} 
where
\begin{align}
\label{cmn1}
&C_{mn}=\frac{1}{2^{m+n+1}}
\sum_{l=0}^m\sum_{k=0}^n {m \choose l}{n \choose k}\frac{(-1)^k+(-1)^l}{ k+l+1}, 
 \end{align}
\vspace{-5ex}

\noindent and
\vspace{-5ex}

\begin{align}
\label{dmn1}
&D_{mn}=\frac{1}{2^{m+n}}
\sum_{l=0}^m\sum_{k=0}^n {m \choose l}{n \choose k} \frac{(-1)^k+(-1)^l}{ k+l+1}(m-l+n-k).
\end{align}
 Using the  formula\cite{garrappa07}
 \begin{align}
\label{garr}
  \sum_{k=0}^n  {n \choose k} \frac{(-1)^k }{k+l+1} = \frac{n!\,l!}{(n+l+1)!}
\end{align}
the summations in Eq. (\ref{cmn1}) can be carried out and one obtains a simple expression for $C_{mn}$, 
 \begin{align}
\label{cmn2}
C_{mn}=\frac{n!\, m!}{(n+m+1)!}.
\end{align}
The  corresponding expression for $D_{mn}$ can be obtained from that  for $C_{mn}$. After  a few  simple manipulations one finds that 
\vspace{-6ex}

 \begin{align}
\label{dmn2}
D_{mn}=2(m+n+1)C_{mn}-\delta_{m0}-\delta_{n0},
\end{align}
\vspace{-6ex}

\noindent
where $\delta_{ij}$ is the Kronecker symbol. 

\section{Proof of Eq. (\ref{hfdif1})}

 To derive Eq.  (\ref{hfdif1}) we have to extract terms proportional to $r^{4N+2}$ that  
appear in the integrals $M_1$ and $M_2$.  To do so, we need an  explicit expression 
for the  remainder $\mathcal{R}_{4N+2}$  in Eq.  (\ref{hfdif1}).    
Representing  $M_1$ in terms of the $I_{mn}$ integrals  and invoking  the asymptotic relation      
(\ref{asymi2}) one  obtains 
 \begin{align}
\label{m1ibis}
\begin{split}
 \mathcal{R}_{4N+2} =& 
f(r)  [ 2\alpha(N+1)I_{2N,2N+1}  -2  Z I_{2N,2N+1} + r^{-1}I_{2N+1,2N+1}  ]\\
+&r^{-1}f'(r)[
(N+2)I_{2N+1,2N+1}+Nr^2I_{2N-1,2N+1}-NI_{2N-1,2N+3}]  +\mathcal{R}_{4N+1}.
\end{split}
\end{align}
We  expand now  the $I_{mn}$ integrals in Eqs.   Eq.  (\ref{hfdif1}) and (\ref{m1ibis}) 
with the help of Eq. (\ref{asymi2})  and after some rearrangements and simplifications 
we arrive  the following formula for $M_1$:
\begin{align}\label{B2}
M_1=r^{4N+2}\,\frac{e^{-2\alpha
r}}{2\alpha}\big[r \,\Xi_{4N+3}+\Omega_{4N+2}+  \mathcal{O}(r^{-1})\big],
\end{align}
where
\begin{align}\label{B2a}
\Xi_{4N+3}=- [f(r)(\alpha^2+E)-2\alpha f'(r )+f''(r) ]C_{2N+1,2N+1}, 
\end{align}
cf. Eq. (\ref{m14n4c}) and (\ref{lem1}), and 
\begin{align}
\begin{split}
\Omega_{4N+2}&=
f(r )\Big[
-\frac{1}{2}(\alpha +\frac{E}{2\alpha})D_{2N+1,2N+1}+2(\alpha N+ \alpha -Z) C_{2N,2N+1} +C_{2N+1,2N+1} \Big] \\
 &-f'(r )
\Big[(N+2)C_{2N+1,2N+1}-\frac{1}{4}D_{2N+2,2N+1}+NC_{2N-1,2N+1}\\
&-NC_{2N-1,2N+3}-\frac{1}{4}D_{2N,2N+1}+\frac{1}{4}D_{2N ,2N+3}
\Big]
-\frac{1}{4\alpha}D_{2N+1,2N+1}f''(r ),
\end{split}
\end{align}
 The expression for $\Omega_{4N+2}$ can be simplified using  Eq. (\ref{dmn2}) 
and the following two identities
  holding  for every $N\geq0$:
\begin{align}\label{lem2}
&2(N+1)C_{2N,2N+1}-\frac{1}{2}(4N+3)C_{2N+1,2N+1}=\frac{1}{2}(2N+3)C_{2N,2N+1},\\[1ex]
&NC_{2N-1,2N+1}-NC_{2N-1,2N+3}-(2N+1)C_{2N,2N+1}+(2N+2)C_{2N,2N+3}=0.
\end{align}
 \noindent
The result of these simplifications is
\begin{align}
\label{xi3}
\begin{split}
\Omega_{4N+2}&=
f(r) \Big[
\frac{\alpha}{2}(2N+3)C_{2N,2N+1}-2ZC_{2N,2N+1}
+C_{2N+1,2N+1}\\
&-\frac{E}{2\alpha}(4N+3)C_{2N+1,2N+1} \Big]-f'(r )C_{2N+1,2N+1}
-\frac{4N+3}{2\alpha}C_{2N+1,2N+1}f''(r ).
\end{split}
\end{align}

 We still need to determine the last required ingredient -- the terms proportional to $r^{4N+2}$ that are  in contained $M_2$.  Expressing $M_2$ in terms of $I_{mn}$ integrals we find
\begin{align}
\begin{split}
M_2&=- f(r )(\alpha^2+E)I_{2N+1,2N}+ \frac{1}{2}\alpha\, r^{-1}\,f'(r )
\Big[
I_{2N+2,2N}+r^2I_{2N,2N}+r^2I_{2N+1,2N-1}\\
&+I_{2N+1,2N+1}-I_{2N,2N+2}-I_{2N+3,2N-1}
\Big]-I_{2N+1,2N}f''(r )
+\mathcal{R}_{4N+1}.
\end{split}
\end{align}
Expansion of every $I_{mn}$ integral according to Eq. (\ref{asymi2}) gives:
\begin{align}\label{B8}
M_2=r^{4N+2}\,\frac{e^{-2\alpha
r }}{2\alpha}\,[\, \Lambda_{4N+2}+\mathcal{O}(r^{-1})\, ],
\end{align}
\vspace{-5ex}

\noindent where
\vspace{-5ex}

 \begin{align}
\label{xi3p}
 \Lambda_{4N+3}&=-C_{2N+1,2N}\left[(\alpha^2+E)f(r )-2\alpha f'(r )+f''(r )\right].
\end{align}
To derive Eq. (\ref{xi3p}) we used the  following relation holding  for every $N\geq0$:
\begin{align}\label{lem3}
C_{2N,2N}+C_{2N+1,2N-1}+C_{2N+1,2N+1}-C_{2N+3,2N-1}=4C_{2N+1,2N}.
\end{align}

   We now have all elements needed to construct the two leading  terms of the r.h.s of Eq.~ (\ref{hf1}).
 Using Eqs.  (\ref{B2}) and (\ref{B8}) one finds that the r.h.s. of Eq.~ (\ref{hf1}) can be written as 
\begin{align}
\label{hf1xi}
c_N^3  \, r^{4N+2}\,\frac{e^{-2\alpha
r }}{2\alpha}\big[c_N\left(r \,\Xi_{4N+3}+\Omega_{4N+2}\right)+4c_{N-1}\, \Lambda _{4N+2}
+\mathcal{O} (r^ {-1} )\big] ,
\end{align}
where $\Xi_{4N+3}$, $\Omega_{4N+2}$ and $\Lambda_{4N+2}$ are given by Eqs. (\ref{B2a}), (\ref{xi3}) and (\ref{xi3p}), respectively, and $c_N$ and $c_{N-1}$ are defined through Eq. (\ref{hfs}). 
The factor of $4$ in  front of $c_{N-1}$ is  a result of the symmetry discussed below Eq. (\ref{matrix1})
and (\ref{matrix2}).  
By neglecting the terms of the
order lower than $r^{4N+2}$ and equating the remaining ones to zero we obtain the required 
differential equation for the function that determines the large-$r$ asymptotic behavior of $f(r)$ 
\begin{align}
\label{asym0}
r\,\Xi_{4N+3}+\Omega_{4N+2}+4b_{N}\,  \Lambda_{4N+2}=0,
\end{align}
where $b_N=c_{N-1}/c_N$.
Inserting into Eq. (\ref{asym0})  the explicit expressions  for $\Xi_{4N+3}$,  $\Omega_{4N+2}$,  and 
$\Lambda_{4N+2}$, given by Eqs. (\ref{B2a}),  (\ref{xi3}), and (\ref{xi3p}),  dividing  by $C_{2N+1,2N+1}$ and   using the  trivial identity:
\begin{align}
\frac{C_{2N,2N+1}}{C_{2N+1,2N+1}}=\frac{4N+3}{2N+1},
\end{align}
one arrives at Eq. (\ref{hfdif1}).


\newpage

\linespread{1.0}
\begin{ruledtabular}
\begin{table}
\caption{The   values of the parameters $B$ and $\rho$ determining the asymptotic behavior of $f(r)$,  Eq. (\ref{asymfull}). Two approaches were used  to fix  the exponent  $\alpha$:   optimization
of the energy obtained with the correlated wave function of  Eq. (\ref{psi1}) and the  ``bare-nucleus" value   $\alpha =Z$.}
\label{table1}
\begin{tabular}{ccD{-}{-}{1.8}cc}\\[-2.5ex]
Z & $\alpha$ & \multicolumn{1}{c}{$E$} & $B$ & $\rho$ \\[0.6ex]
\hline\\[-2ex]
\multicolumn{5}{c}{energy optimized $\alpha$} \\[0.6ex]
\hline
1 & 0.84267 & -0.509378 & 0.128966 & 0.322138 \\
2 & 1.84833 & -2.891254 & 0.147959 & 0.147577 \\
3 & 2.85039 & -7.268487 & 0.154375 & 0.095543 \\
4 & 3.85144 & -13.64459 & 0.157585 & 0.070615 \\
5 & 4.85208 & -22.02025 & 0.159510 & 0.055997 \\
6 & 5.85251 & -32.39568 & 0.160792 & 0.046391 \\
7 & 6.85282 & -44.77099 & 0.161707 & 0.039598 \\
8 & 7.85305 & -59.14622 & 0.162393 & 0.034540 \\
\hline\\[-2ex]
\multicolumn{5}{c}{$\alpha=Z$} \\[0.6ex]
\hline
1 & 1.00000 & -0.498452 & 0.293989 & 0.124612 \\
2 & 2.00000 & -2.879363 & 0.303131 & 0.062623 \\
3 & 3.00000 & -7.256353 & 0.306238 & 0.041754 \\
4 & 4.00000 & -13.63235 & 0.307799 & 0.031309 \\
5 & 5.00000 & -22.00795 & 0.308737 & 0.025041 \\
6 & 6.00000 & -32.38335 & 0.309363 & 0.020864 \\
7 & 7.00000 & -44.75863 & 0.309811 & 0.017880 \\
8 & 8.00000 & -59.13384 & 0.310147 & 0.015643 \\
\end{tabular} 
\end{table} 
\end{ruledtabular}

\clearpage

\begin{ruledtabular}
\begin{table}
\caption{Ground-state energies of the helium atom obtained with approximate wave functions of  Eqs. (\ref{psi1}) and (\ref{kutz1}).   
Results obtained with the linear,  $1+r/2$, and exponential, $(1+2\gamma -e^{-\gamma r})/(2\gamma)$,  
correlation factors  are denoted by R12 and F12, respectively. The parameter $\gamma$\! =\! 0.2  is close to optimal. 
Eqs.~(\ref{rs1}) and  (\ref{rs2}) are evaluated with $n=0$.  The orbital exponent $\alpha$ was always set 
equal to 1.84833.}
\label{table2}
\begin{tabular}{cD{.}{.}{1.5}D{.}{.}{1.5}}\\[-2ex]
   $f(r )$          & \multicolumn{1}{c}{\mbox{wave function of Eq. (\ref{psi1})}} & \multicolumn{1}{c}{\mbox{wave of
function Eq.
(\ref{kutz1})}} \\[1ex]
\hline\\[-2ex]
R12       & -2.887447 & -2.903014 \\
F12 ($\gamma$=$0.5$)       & -2.886746 & -2.902976 \\
F12 ($\gamma$=$1.0$)       & -2.874472 & -2.900928 \\
F12 ($\gamma$=$0.2 $)& -2.890349 & -2.903277 \\
Eq.~(\ref{rs1})       & -2.890886 & -2.903266 \\
Eq.~(\ref{rs2})      & -2.891048 & -2.903325 \\
limit                       & -2.891254^a & -2.903512^b  \\ 
\end{tabular} 
\begin{flushleft}
$^a${\small obtained by numerical integration of differential  equation}\\
$^b${\small obtained by expanding  $f(r)$ in powers of $r$ (saturated results, all digits shown are correct)}.
\end{flushleft}
\end{table} 
\end{ruledtabular}
\clearpage
\begin{figure}
\caption{The correlation factor $f(r)$ calculated for the helium atom  
using the wave function  of Eq.  (\ref{psi1}) and
$\alpha$=1.84833. Red solid line is the result of   numerical propagation of
Eq.~(\ref{dif1s1s}). Black dash-dotted line is the variational solution with
$f(r)$ expanded in the powers of $r$.
Green dashed line is the  first term of the asymptotic expansion of $f(r)$.
Blue dotted line is used for the short-range factor $1+\frac{1}{2}r$.}
\vspace{0.5cm}
\label{fig1s}
\includegraphics[width=0.8\textwidth]{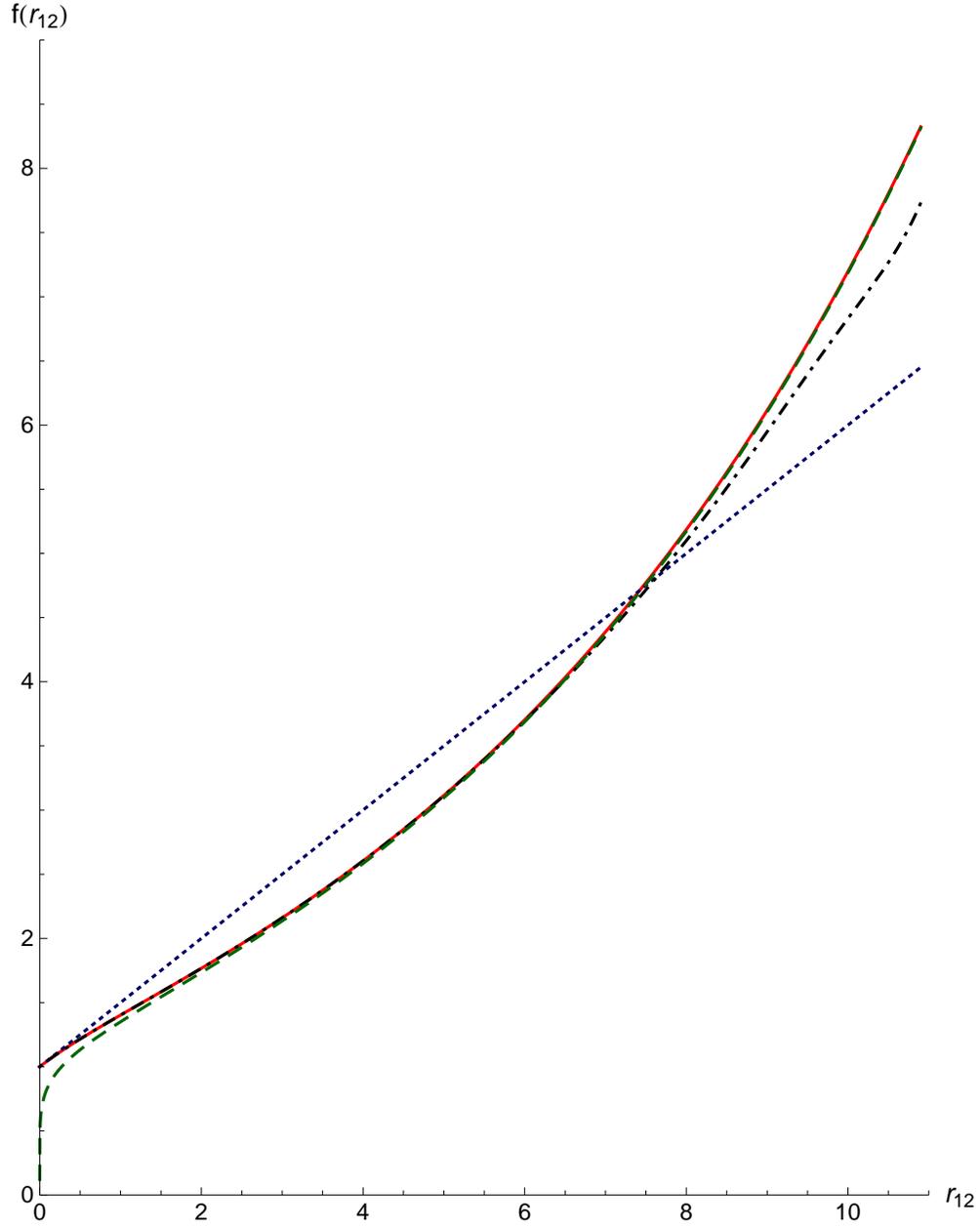}
\end{figure}
\clearpage
\begin{figure}
\caption{The correlation factor $f(r )$ calculated for the helium atom  
using the wave function  of Eq.~(\ref{psi2})
 with $\alpha$=0.321454 and $\beta$=1.968451. 
The explanation of lines is the same as in Fig. 1, except that the short-range
correlation factor, marked by the blue dotted line, in now 
 $1+\frac{1}{4}r$.}
\vspace{0.5cm}
\label{figt}
\includegraphics[width=0.8\textwidth]{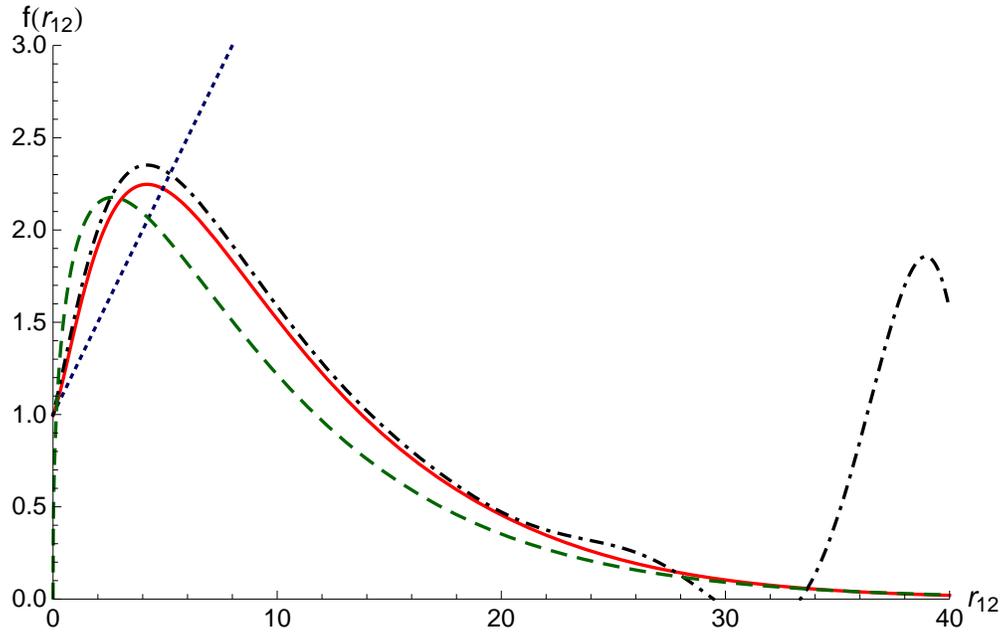}
\end{figure}
\clearpage
\begin{figure}
\caption{The correlation factor $f(r )$ calculated for the helium atom by
using the Gaussian wave function of Eq.~(\ref{psi4}) and
$\alpha=0.8598$. Red solid line is the result of  numerical solution  of
the differential equation
(\ref{1s1sgauss}). Black doted-dashed line is the variational solution with
$f(r)$ expanded in the powers of $r$.
Green dashed line is the leading term of the asymptotic expansion of $f(r)$
calculated for the relevant values of
  parameters. Blue dotted line ($1+\frac{1}{2}r$) is plotted for the
comparison purposes. Two different plot
ranges are given separately to improve  the readability.}
\vspace{0.5cm}
\label{figgauss}
\includegraphics[width=0.8\textwidth]{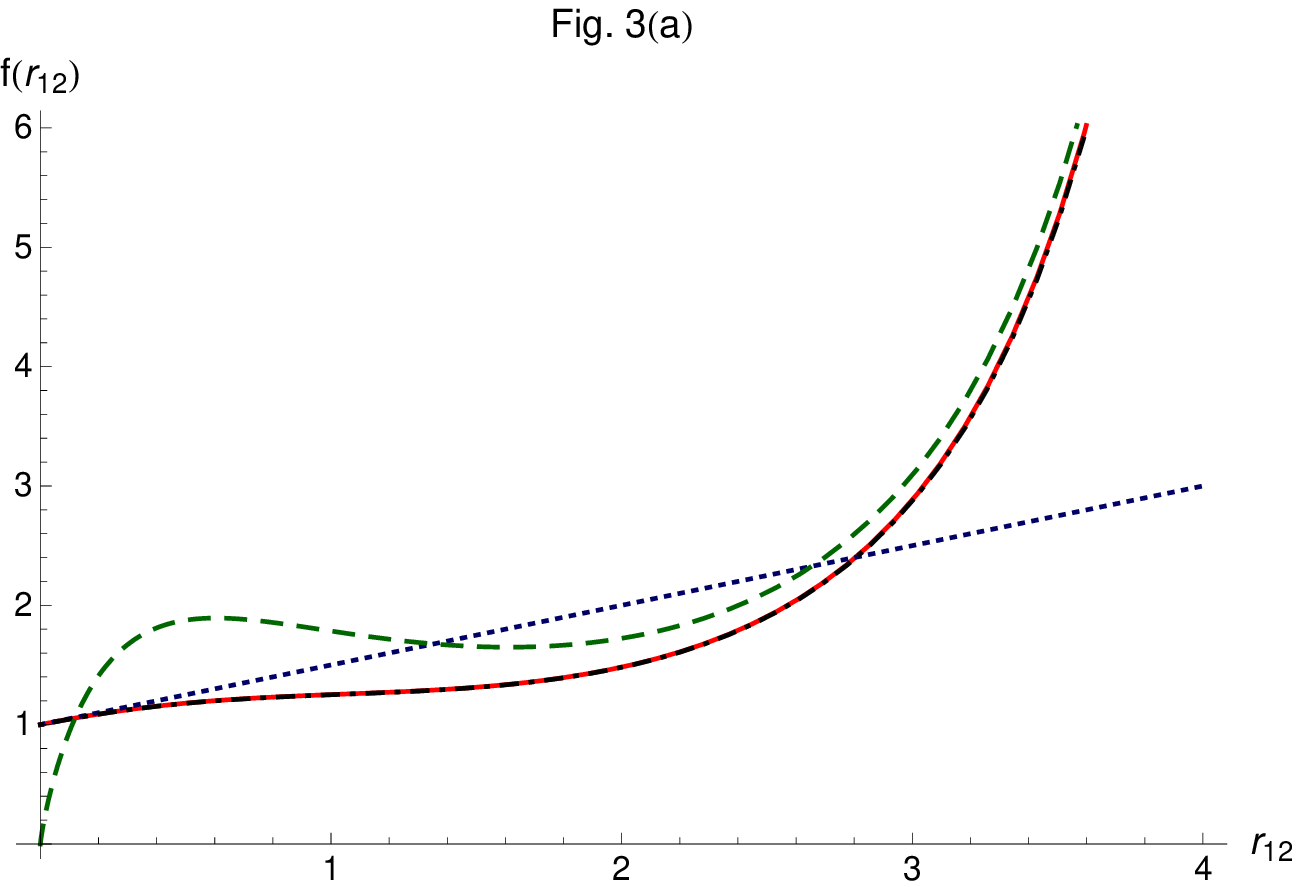}\\
\vspace{1.5cm}
\includegraphics[width=0.8\textwidth]{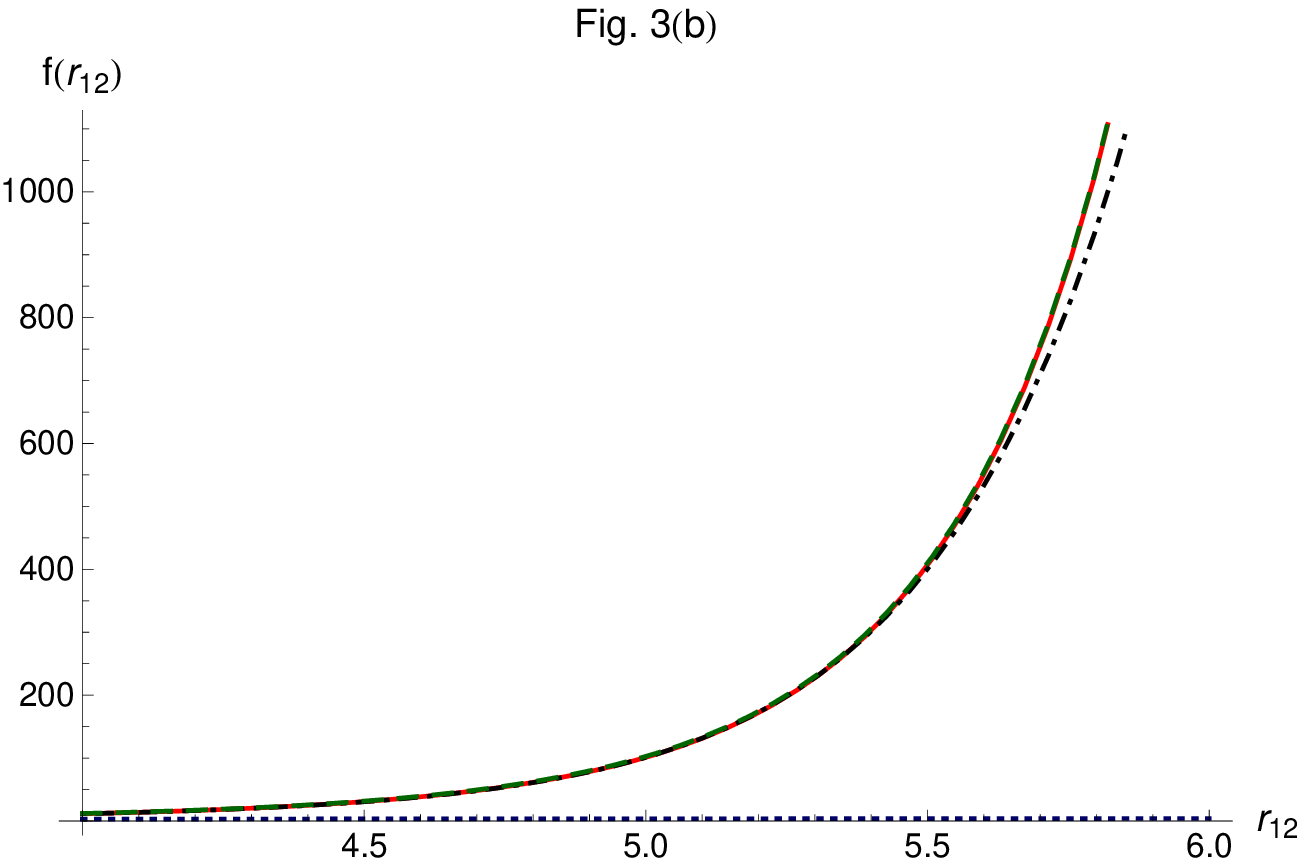}
\end{figure}
\clearpage
\begin{figure}
\caption{Plot of $\rho(N)$ parameter calculated for the helium atom [black
curve, Eq. (\ref{asym1srho})]
compared to its  large-$N$  asymptote [red line, Eq. (\ref{skewrho})]. The
corresponding curves for the other helium-like ions
were not included since they are barely distinguishable with the adopted scale
of the plot.}
\vspace{0.5cm}
\label{figrho}
\includegraphics[width=0.8\textwidth]{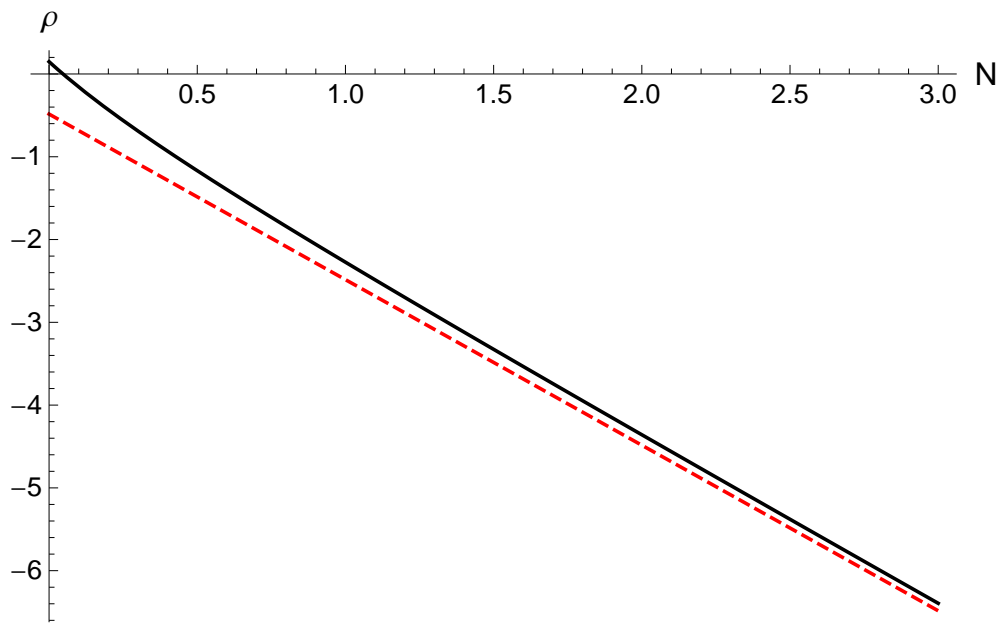}
\end{figure}
\clearpage
\begin{figure}
\caption{The correlation factor $f(r)$ calculated for the helium atom by
using the wave function (\ref{hff12})
($N=2$) and $\alpha=1.920904$. Red solid line is the result of the numerical
propagation of the
corresponding differential equation. Black dotted line is the variational
solution with $f(r)$ expanded in the
powers of $r$. Green dashed line is the leading term of the asymptotic
expansion of $f(r)$ calculated for the
relevant values of the parameters [see Eqs. (\ref{bianka}) and (\ref{rhohf})].
Blue dotted line
($1+\frac{1}{2}r$) is plotted for the comparison purposes.}
\vspace{0.5cm}
\label{fighf}
\includegraphics[width=0.8\textwidth]{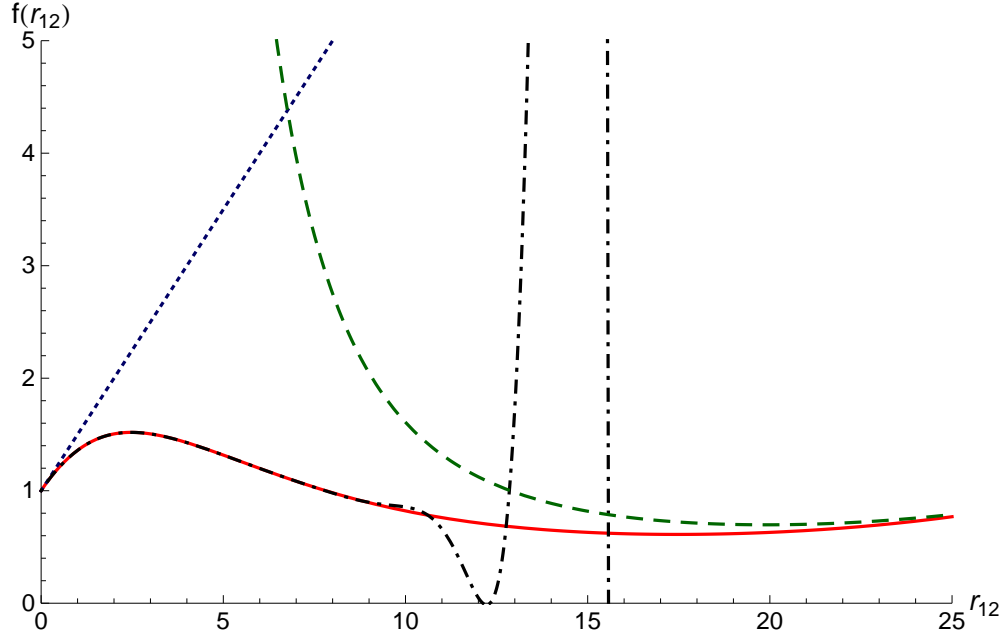}
\end{figure}

\end{document}